\documentclass[11pt]{article}
\usepackage{epsfig}
\usepackage{amscd}
\usepackage{amssymb}
\usepackage{amsfonts}
\usepackage{amsmath}
\usepackage{pstricks}
\usepackage{pst-node}
\usepackage{mathrsfs}
\usepackage{graphics,epsfig,psfrag}

\newcommand{\ssc}{\scriptscriptstyle}
\newcommand{\be}{\begin{equation}}
\newcommand{\ee}{\end{equation}}
\newcommand{\bea}{\begin{eqnarray}}
\newcommand{\eea}{\end{eqnarray}}

\begin{document}

\makeatletter
\@addtoreset{equation}{section}
\makeatother
\renewcommand{\theequation}{\thesection.\arabic{equation}}

\begin{titlepage}

\baselineskip =15.5pt
\pagestyle{plain}
\setcounter{page}{0}

\begin{flushright}
\end{flushright}

\vfil

\begin{center}
{\huge {\bf Geometric Dequantization}}
\end{center}

\vfil

\begin{center}
{\large A. A. Abrikosov, Jr.}\\
\vspace {1mm}
ITEP, Bol. Cheremushkinskaya, 25;
        Moscow, 117259 Russia \\
        e-mail: {\it persik@itep.ru}
\vspace {1mm}

\bigskip

{\large E. Gozzi and D. Mauro}\\
\vspace {1mm}
Department of Theoretical Physics, University of Trieste \\
Strada Costiera 11, Miramare-Grignano 34014, Trieste\\ and INFN,  
Sezione di Trieste, Italy\\
e-mail: {\it gozzi@ts.infn.it} and {\it mauro@ts.infn.it}
\vspace {1mm}
\vspace{3mm}
\end{center}

\vfil

\begin{abstract}
\noindent Dequantization is a set of rules which turn quantum mechanics (QM) into classical mechanics (CM). It is not the WKB limit of QM. In this paper we show that, by extending time to a 
3-dimensional ``supertime", we can dequantize the system in the sense of turning the Feynman path integral version of QM into the functional counterpart of the Koopman-von Neumann operatorial approach to CM. Somehow this procedure is the inverse of geometric quantization and we present it in three different polarizations: the Schr\"odinger, the momentum and the coherent states ones. 
\end{abstract}
\vfil
\end{titlepage}

\newpage 
\section{Introduction}

Quantum mechanics has turned out to be the best tested theory invented by men. This is quite a surprise for us because, somehow, QM was built by a series of ``trials and errors" attempts starting from N. Bohr orbits and ending with the quantization rules proposed by Dirac and Heisenberg. These quantization rules, which we could summarize as follows
\begin{displaymath}
\begin{array}{c}
\displaystyle p\, \longrightarrow \, -i\hbar \frac{\partial}{\partial q}, \medskip \\
\displaystyle \{ \quad , \quad \} \, \longrightarrow \, -\frac{i}{\hbar}[\quad , \quad], 
\end{array}
\end{displaymath}
are totally unexplained: they just work but nobody understands why. At the same time QM presents a set of phenomena (from tunnelling to teleportation etc.) which cannot be visualized by our macroscopic senses. We feel that both these aspects, the unexplainable quantization rules and the counterintuitive phenomena of QM, could be the signal of some hidden {\it geometry}, difficult 
to visualize. Geometry may be important for QM also for another reason. In fact, despite 
its great experimental success, QM seems not to get along well with gravity, which is the queen of ``{\it geometrical}" theories. The problem is that Einstein gravity, once quantized, does not turn out to be a renormalizable theory. The strategy that physicists have adopted to solve this 
problem is to modify the classical gravity theory of Einstein by adding to it 
extra fields and interactions, like in supergravity, or infinite terms like in string theory.
They never thought of attacking the other horn of the quantum gravity problem that is QM.  ``Attacking", for us, does not mean introducing modifications to QM, but only trying to understand QM from a more ``{\it geometrical}" point of view. This seems to us a reasonable approach if we want to marry QM with the queen of geometrical theories, which is gravity. This reason, together with the one given before, is what triggered the present investigation. 

QM has always been studied at the Hilbert spaces and operators level and not much 
at the space-time level, which we consider the true geometrical level. 
The only approach in this sense has been pioneered by Feynman in the well-known paper \cite{uno} on path integrals which carries significantly the title
``{\it A space-time approach to QM}". In this paper we will continue in this direction by giving a path integral approach \cite{due}
even to classical mechanics (CM). This can be done once CM is formulated in an operatorial form like 
Koopman and von Neumann \cite{tre} did in the Thirties. From now on we will indicate this
path integral formulation of {\it classical} mechanics with the acronym ``{\it CPI}" 
({\it C}lassical {\it P}ath {\it I}ntegral) while we will use for the Feynman {\it P}ath {\it I}ntegral of {\it Q}uantum mechanics the symbol ``{\it QPI}". The operatorial approach to CM developed by {\it K}oopman and {\it v}on {\it N}eumann will be instead indicated with the acronym {\it KvN}.

The CPI can be put in a very compact form by extending time to a 
3-dimensional supertime made of $t$ and of two 
Grassmann partners $\theta, \bar{\theta}$. Quantization, which means a manner to go from the CPI to the QPI, is achieved by properly freezing to zero the two Grassmann partners
of time $(\theta, \bar{\theta})$. This procedure of quantization via a contraction from 3-dim to 1-dim, resembles very much the mechanisms used nowadays in grand-unification
theories where one starts from an 11-dim theory and brings it down, via dimensional reduction,
to 4-dim. It is amazing that quantization itself can be obtained via a dimensional reduction procedure. Of course, it is a dimensional reduction that involves Grassmann variables and so all the geometry
that one gets is very hard to visualize. This ``hidden, hard-to-visualize" {\it geometry}
may be the structure that, as we said at the beginning of this section, is at the origin of all
the ``strange" phenomena of QM. 
Anyhow, much more direct is the procedure of ``{\it dequantization}", which means a manner to go from the QPI to the CPI. This can be achieved via an extension of time $t$
to supertime $(t,\theta,\bar{\theta})$ and of phase space to a superphase space, which appears
naturally \cite{due} in the path integral approach to CM. We call ``{\it geometrical}" this procedure
of dequantization because it involves geometrical structures like time or its Grassmann
partners. At the relativistic field theory level we expect that the role played by time will 
be taken by space-time and this is another reason for the use of the word ``{\it geometrical}" in our procedure. 

The paper is organized as follows. In Sec. 2 we present a brief review of the KvN operatorial 
approach to CM \cite{tre} and of its path integral counterpart (CPI) \cite{due}, with its 
rich geometrical content \cite{quattro}. We do this brief review for completeness and in order to make the paper self-contained. In Sec. 3 we introduce the concept of supertime 
and of superphase space variables and explore some interesting issues related to them. These tools are then used in Sec. 4 where we prove that the path integral weights entering the CPI and the QPI belong to the same multiplet of a universal supersymmetry 
\cite{quattro} present in the CPI. This observation is the starting point in order to understand how the QPI can be turned into the CPI (dequantization procedure). This is realized by keeping the same functional form of the QPI and extending time to supertime and phase space to superphase space. This dequantization procedure is studied in Sec. 4 in the Schr\"odinger and the momentum polarization while in Appendix E is extended to the coherent states basis. In Sec. 4 we also show that the inverse \cite{cinque} of our dequantization procedure resembles very much {\it geometric quantization} \cite{sei}. In Sec. 5 we study the Dyson-Schwinger equation for the QPI and the CPI and show that the rules to pass from one to the other are the same as above i.e., replace time with supertime and phase space with superphase space. In Sec. 6 we formulate more precisely the conditions under which the above dequantization rules can be applied. They cannot be used in any equation or expression but only within path integral expressions. A set of examples is worked out in details. In the same section we show how our dequantization rules have to be applied to QM observables. The conclusions of this paper are contained in Sec. 7 where, besides summarizing what we have done, we put forward some lines of future research. All the calculational details of this paper are confined into several appendices.

\section{Brief Review of the KvN Formalism and of the CPI}

In the Thirties Koopman and von Neumann \cite{tre}, triggered most probably 
by the operatorial aspects of QM, proposed an operatorial formulation
for CM. Their work is basically an extension of the one of Liouville 
who, for classical statistical mechanics, had found the equation of evolution 
for the probability distribution $\rho(p,q)$ defined on phase space. 
The Liouville equation is:
\begin{equation}
\displaystyle i\frac{\partial}{\partial t}\rho =\hat{L}\rho \label{8-1}
\end{equation}
where $\hat{L}$ is the Liouvillian defined as 
\begin{displaymath}
\displaystyle \hat{L}=-i\partial_pH\partial_q+i\partial_qH\partial_p 
\end{displaymath}
with $H$ the Hamiltonian function of the system. We will use
the following notation: ${\cal M}$ indicates the $2n$ dimensional phase space, while $\varphi^a
=(q^1 \cdots q^n,p^1 \cdots p^n)$ with $a=1,\cdots, 2n$ indicate the $2n$ phase
space coordinates. Liouville's probability distributions $\rho(\varphi)$  
are $L^1$ functions because the only request made on them is to be integrable
\begin{displaymath}
\displaystyle \int d \varphi \; \rho(\varphi)=1
\end{displaymath}
where by ``$d\varphi$" we mean ``$d^{2n}\varphi$".
Koopman and von Neumann, instead of using the probability distribution
$\rho(\varphi)$, introduced a {\it Hilbert space} of $L^2$ functions
$\psi(\varphi)$. Furthermore they imposed the following two postulates for $\psi(\varphi)$:

{\bf 1st postulate}: The equation of evolution for $\psi$ is
\begin{equation}
i\frac{\partial\psi}{\partial t}=\hat{L}\psi, \label{9-2}
\end{equation}
that is the same as the Liouville equation (\ref{8-1}). 

{\bf 2nd postulate}: The probability distribution $\rho(\varphi)$ can be built out of the $\psi$ as follows:
\begin{equation}
\rho(\varphi)=|\psi(\varphi)|^2. \label{10-1}
\end{equation}
These two postulates are not in contradiction with the Liouville equation
(\ref{8-1}). In fact, since $\hat{L}$ is a differential operator, linear in 
the first order derivatives, from (\ref{9-2}) and (\ref{10-1}) it is easy to
derive (\ref{8-1}), that means that the same equation is satisfied by both
$\rho$ and $\psi$. This would not happen in QM, where the KvN equation
(\ref{9-2}) is replaced by the Schr\"odinger one and the quantum equation
for $\rho$ is a continuity equation, different from the Schr\"odinger one. This is due to
the presence of second order derivatives in the Schr\"odinger operator.

In \cite{sette} we called the $\psi$ of (\ref{9-2}) ``{\it the KvN waves}" and
we thoroughly analyzed their physics. The overall KvN formalism 
is formally very similar to the one
of QM, even if the Liouvillian $\hat{L}$ \cite{sette}
reproduces the {\it classical} dynamics. Like every operatorial formalism, also
the KvN one has a path integral counterpart, which has been fully developed in Ref.
\cite{due} and reproduces the
kernel of evolution associated to the classical equation (\ref{9-2}). In CM if we ask which is the transition amplitude $K(\varphi^a,t|
\varphi^a_{\ssc 0},t_{\ssc 0})$ of being in $\varphi^a$ at time $t$ having started from $\varphi^a_{\ssc 0}$ at time $t_{\ssc 0}$, the answer is 
\begin{equation}
\displaystyle K(\varphi^a,t|\varphi^a_{\ssc 0},
t_{\ssc 0})=\delta[\varphi^a-\phi^a_{\textrm{cl}}(t;\varphi_{\ssc 0},t_{\ssc 0})]  \label{11-1}
\end{equation}
where $\phi^a_{\textrm{cl}}(t;\varphi_{\ssc 0},t_{\ssc 0})$ is the solution of Hamilton's equations with initial conditions $\varphi_{\ssc 0}$ 
\begin{equation}
\displaystyle \dot{\varphi}^a=\omega^{ab}\frac{\partial H}{\partial \varphi^b}
\label{12-1} 
\end{equation}
and $\omega^{ab}$ indicates the symplectic matrix \cite{otto}.
We can rewrite (\ref{11-1}) as a sum over all possible intermediate 
configurations, after having sliced the interval $(t-t_{\ssc 0})$ in $N$ intervals
\begin{eqnarray}
K(\varphi^a,t|\varphi_{\ssc 0}^a,t_{\ssc 0})\hspace{-0.4cm}&=& \hspace{-0.4cm} \sum_{\varphi_i}K(\varphi^a,t|\varphi_{\ssc N-1},t_{\ssc N-1})
K(\varphi_{\ssc N-1},t_{\ssc N-1}|\varphi_{\ssc N-2},t_{\ssc N-2})\cdots K(\varphi_{\ssc 1},t_{\ssc 1}|\varphi^a_{\ssc 0},t_{\ssc 0})\nonumber\\
\hspace{-0.4cm}&=&\hspace{-0.4cm} \prod_{j=1}^N\int d\varphi_j \,\delta[\varphi^a_j-\phi^a_{\textrm{cl}}
(t_j;\varphi_{j-1},t_{j-1})]  \label{12-2} \\
&\xrightarrow{N\rightarrow\infty}& \int {\mathscr D}\varphi \,
\widetilde{\delta}
[\varphi^a-\phi_{\textrm{cl}}^a(t;\varphi_{\ssc 0},t_{\ssc 0})]. \nonumber
\end{eqnarray}
In the first line of (\ref{12-2}) $|\varphi_i,t_i)$ denotes a set of intermediate configurations
between $|\varphi_{\ssc 0}^a,t_{\ssc 0})$ and $|\varphi^a,t)$, while in the last one 
we have sent the number of slices 
to infinity recovering in this manner a functional integration. 
The $\widetilde{\delta}[\qquad]$ indicates a functional Dirac delta which effectively gives 
weight ``one" to the classical trajectory and weight ``zero" to all the other ones. 
This functional delta can be rewritten as a delta on the Hamilton equations 
of motion (\ref{12-1}) via the introduction of a suitable functional determinant 
\begin{equation}
\displaystyle \widetilde{\delta}[\varphi^a-\phi_{cl}^a(t;\varphi_{\ssc 0},t_{\ssc 0})]=
\widetilde{\delta} [\dot{\varphi}^a-\omega^{ab}\partial_bH]\textrm{det}(
\delta_b^a\partial_t-\omega^{ac}\partial_c\partial_bH). \label{13-1}
\end{equation}
Let us now perform a Fourier transform of the Dirac delta on the RHS of (\ref{13-1}) 
introducing $2n$ extra variables $\lambda_a$ and let us exponentiate the determinant
using $4n$ Grassmann variables $c^a$, $\bar{c}_a$. The final result is the
following one:
\begin{equation}
\displaystyle K(\varphi^a,t|\varphi^a_{\ssc 0},t_{\ssc 0})=\int {\mathscr D}^{\prime\prime}
\varphi {\mathscr D}\lambda {\mathscr D} c {\mathscr D}\bar{c}
\; \textrm{exp} \left[i\int_{t_{\ssc 0}}^{t} d\tau \widetilde{\cal L}\right]
\label{13-2}
\end{equation} 
where ${\mathscr D}^{\prime\prime}\varphi$ indicates that the integration 
in $\varphi$ is over the paths $\varphi(t)$ with fixed end points $\varphi_{\ssc 0}$ and
$\varphi$, while the other integrations ${\mathscr D}\lambda {\mathscr D}c{\mathscr
D}\bar{c}$ include the end points of the paths in $\lambda$, $c$, $\bar{c}$.
The $\widetilde{\cal L}$ which appears in (\ref{13-2}) is
\begin{equation}
\displaystyle \widetilde{\cal L}=\lambda_a\dot{\varphi}^a+i\bar{c}_a\dot{c}^a
-\lambda_a\omega^{ab}\partial_bH-i\bar{c}_a\omega^{ad}\partial_d\partial_bHc^b
\label{14-1a}
\end{equation}
and its associated Hamiltonian is:
\begin{equation}
\widetilde{\cal H}=\lambda_a\omega^{ab}\partial_bH+i\bar{c}_a\omega^{ad}
\partial_d\partial_bHc^b \label{14-1b}.
\end{equation}
The path integral (\ref{13-2}) has an operatorial counterpart and we shall now check
that this is the KvN theory. Let us first remember \cite{uno} that from the kinetic
part of the Lagrangian appearing in the weight of a path integral, one can deduce the 
commutators of the associated operator theory. In our case from (\ref{14-1a}) we get the following graded commutators:
\begin{equation}
[\hat{\varphi}^a,\hat{\lambda}_b]=i\delta_b^a, \qquad \qquad
[\hat{c}^a,\hat{\bar{c}}_b]=\delta_b^a \label{14-2}
\end{equation}
while all the others are zero, in particular $[\hat{p},\hat{q}]=0$ and this is the
clear indication that what we have obtained is CM and not QM. Despite having objects that do not commute, as indicated by (\ref{14-2}), in the CPI there are no ordering problems. This issue is analyzed in detail in Appendix G.

The commutators (\ref{14-2}) can be realized in various ways. One way is to implement
$\hat{\varphi}^a$ and $\hat{c}^a$ as multiplicative operators and $\hat{\lambda}_a$ and $\hat{\bar{c}}_b$ as derivative ones:
\begin{equation}
\displaystyle \hat{\lambda}_a=-i\frac{\partial}{\partial\varphi^a}, \qquad
\hat{\bar{c}}_b=\frac{\partial}{\partial c^b} \label{14-3}.
\end{equation}
This realization automatically defines a basis $|\varphi^a,c^a\rangle$ given by:
\begin{equation}
\hat{\varphi}^a|\varphi^a,c^a\rangle =\varphi^a|\varphi^a,c^a\rangle, \qquad 
\hat{c}^a|\varphi^a,c^a\rangle=c^a|\varphi^a,c^a\rangle. \label{14-4}
\end{equation}
The realization (\ref{14-3}) is not the only possible one. Among the $4n$ operators 
$\hat{\varphi}^a\equiv (\hat{q},\hat{p})$ and 
$\hat{\lambda}_a\equiv (\hat{\lambda}_q,\hat{\lambda}_p)$ we could have chosen
to implement $\hat{q}$ and $\hat{\lambda}_p$ as multiplicative operators
and, as they commute because of the commutators (\ref{14-2}), diagonalize them simultaneously. $\hat{p}$ and $\hat{\lambda}_q$ must then be realized as derivative operators:
\begin{displaymath}
\displaystyle \hat{p}=i\frac{\partial}{\partial \lambda_p}, \qquad \quad
\displaystyle \hat{\lambda}_q=-i\frac{\partial}{\partial q}. \label{15-1}
\end{displaymath}
The same can be done for $\hat{c}^a\equiv (\hat{c}^q,\hat{c}^p)$ and $\hat{\bar{c}}_a
\equiv (\hat{\bar{c}}_q,\hat{\bar{c}}_p)$. The basis defined by this overall
realization is 
\begin{equation}
\begin{array}{l}
\hat{q}|q,\lambda_p,c^q,\bar{c}_p\rangle
=q|q,\lambda_p,c^q,\bar{c}_p\rangle \medskip\\
\hat{\lambda}_p|q,\lambda_p,c^q,\bar{c}_p\rangle=
\lambda_p|q,\lambda_p,c^q,\bar{c}_p\rangle \medskip\\
\hat{c}^q|q,\lambda_p,c^q,\bar{c}_p\rangle=
c^q|q,\lambda_p,c^q,\bar{c}_p\rangle \medskip \\
\hat{\bar{c}}_p|q,\lambda_p,c^q,\bar{c}_p\rangle=
\bar{c}_p|q,\lambda_p,c^q,\bar{c}_p\rangle. \label{15-2}
\end{array}
\end{equation}
Let us for the moment stick to the realization (\ref{14-3})-(\ref{14-4}) and build
the {\it operator} associated to the Hamiltonian (\ref{14-1b}). If we restrict
ourselves to its non-Grassmann part, which
we will indicate with $\widetilde{\cal H}_{\ssc B}$ (B for Bosonic), we get:
\begin{displaymath}
\displaystyle \widetilde{\cal H}_{\ssc B}=\lambda_a\omega^{ab}\partial_bH\, \longrightarrow
\, \hat{\widetilde{\cal H}}_{\ssc B}=-i\omega^{ab}\partial_bH\partial_a=-i\partial_p
H\partial_q+i\partial_qH\partial_p=\hat{L}.
\end{displaymath}
This tells us that $\hat{\widetilde{\cal H}}_{\scriptscriptstyle B}$ is exactly the Liouville operator, so the kernel (\ref{13-2}), in its bosonic part, can be written as:
\begin{equation}
\displaystyle K(\varphi, t|\varphi_{\ssc 0},t_{\ssc 0})=\langle \varphi|\exp
-i\hat{L}(t-t_{\ssc 0})|\varphi_{\ssc 0} \rangle \label{16-2}.
\end{equation} 
This confirms (modulo the Grassmann part) that the path integral
(\ref{13-2}) is exactly the functional counterpart of the KvN theory.

What can we say about the Grassmann part of $\widetilde{\cal H}$? This has been thoroughly 
analyzed in Refs. \cite{due} and \cite{quattro}. Here we will be rather brief,
referring the reader to papers \cite{due}\cite{quattro} for further details. Once
the operatorial realization (\ref{14-3}) is used, the Hamiltonian $\widetilde{\cal H}$
in (\ref{14-1b}) is turned into the following operator:
\begin{equation}
\displaystyle \hat{\widetilde{\cal H}}=-i\omega^{ab}\partial_bH\partial_a
-i\omega^{ab}\partial_b\partial_dH\hat{c}^d\frac{\partial}{\partial c^a}.
\label{16-3}
\end{equation}
Via the commutators (\ref{14-2}) the equations of motion of the $\hat{c}^a$ turn out
to be
\begin{displaymath}
\dot{\hat{c}}^a=i[\widetilde{\cal H}, \hat{c}^a]=\omega^{ab}\partial_b\partial_dH\hat{c}^d. \label{17-1}
\end{displaymath}
Note that these are the same equations of motion of the differentials $d\hat{\varphi}^a$,
which can be obtained by doing the first variation of the Hamilton equations
(\ref{12-1}). So we can make the identification 
\be
\displaystyle \hat{c}^a=d\hat{\varphi}^a, \label{17-2}
\ee
that means the $\hat{c}^a$ are essentially a basis for the differential forms of the phase space \cite{otto}. In general, a form has the expression 
\be
\displaystyle F(\hat{\varphi}, d\hat{\varphi})=
F_{\ssc 0}(\hat{\varphi})+F_a(\hat{\varphi})d\hat{\varphi}^a+F_{ab}(\hat{\varphi})d
\hat{\varphi}^a\wedge d\hat{\varphi}^b+\cdots, \label{17-3}
\ee
where ``$\wedge$" is the antisymmetric wedge product usually defined among forms
\cite{otto}. Via the identification (\ref{17-2}) the form (\ref{17-3}) becomes a 
function of $\hat{c}$ 
\be
\displaystyle F(\hat{\varphi}, d\hat{\varphi})\,\longrightarrow \,
F(\hat{\varphi}, \hat{c})=F_{\ssc 0}(\hat{\varphi})+F_a(\hat{\varphi})\hat{c}^a+F_{ab}(\hat{\varphi})
\hat{c}^a\hat{c}^b+\cdots \label{17-4}
\ee
where there is no need to introduce the wedge product anymore, because this is
automatically taken care of by the Grassmann character of the operators $\hat{c}^a$. Like $\hat{c}^a$ can be given a geometrical interpretation, the same can be done for the 
$\hat{\widetilde{\cal H}}$ in (\ref{16-3}). This turns out \cite{due}\cite{quattro}
to be nothing else than the Lie derivative along the Hamiltonian vector field 
\cite{otto} associated to $H$. So this is basically the geometrical object which extends
the Liouville operator to the space of forms. While the Liouville operator
makes the evolution only of the zero forms that are the $F_{\ssc 0}(\hat{\varphi})$ of (\ref{17-4}), the Lie derivative $\hat{\widetilde{\cal H}}$ makes the evolution of
the entire $F(\hat{\varphi},\hat{c})$ with all its components appearing on the 
RHS of (\ref{17-4}). Besides the forms and the Lie derivatives, also many other
geometrical structures, like Lie brackets, exterior derivatives, etc. \cite{otto}
can be written in the language of our CPI and this has been done in great details 
in Refs. \cite{due} and \cite{quattro}. 

Having now these extra variables $c^a$, the ``Koopman-von Neumann waves"
$\psi(\varphi)$ will be extended to functions of both $\varphi$ and $c$: $\psi(\varphi,c)$.
Of course, in order to have a Hilbert space, a proper scalar product 
among these extended waves must be given. All this has been done in detail in Ref. \cite{sette}.
Using the basis defined in (\ref{14-4}) and a proper scalar product we could represent
the ``KvN waves" $\psi(\varphi,c)$ as:
\be
\displaystyle \langle \varphi^a,c^a|\psi\rangle =\psi(\varphi^a,c^a). \label{19-1}
\ee
In this basis we could also generalize the kernel of propagation (\ref{16-2}) 
to the following one 
\be
\displaystyle K(\varphi,c,t|\varphi_{\ssc 0},c_{\ssc 0},t_{\ssc 0})
\equiv \langle \varphi,c,t|\varphi_{\ssc 0},c_{\ssc 0},t_{\ssc 0}\rangle=\langle \varphi,c|
\exp -i\hat{\widetilde{\cal H}}(t-t_{\ssc 0})|\varphi_{\ssc 0},c_{\ssc 0}\rangle, \label{19-2}
\ee
whose path integral representation is
\be
\displaystyle K(\varphi,c,t|\varphi_{\ssc 0},c_{\ssc 0},t_{\ssc 0})
=\int {\mathscr D}^{\prime\prime} \varphi {\mathscr D}\lambda
{\mathscr D}^{\prime\prime}c{\mathscr D}\bar{c} \, \exp i\int_{t_{\ssc 0}}^t 
d\tau \widetilde{\cal L} \label{19-3}.
\ee
The difference with respect to (\ref{13-2}) is in the measure of integration,
which in (\ref{19-3}) has the initial and final $c$ {\it not} integrated over.

\section{Supertime and Superphase space}

We have seen in the previous chapter how the $8n$ variables $(\varphi^a,\lambda_a,c^a,\bar{c}_a)$,
which enter the Lagrangian (\ref{14-1a}), can be turned into operators by the path integral 
(\ref{13-2}). Actually if one looks at $\widetilde{\cal L}$ and $\widetilde{\cal H}$ not as weight
of a path integral but as a standard Lagrangian and Hamiltonian of a classical system, one could then look
at $(\varphi^a,\lambda_a,c^a,\bar{c}_a)$ as coordinates of an {\it extended phase space}.
The variation of the action $\int d\tau \widetilde{\cal L}$ with respect to $\lambda_a$,
$\bar{c}_a$, $c^a$, $\varphi^a$ would give respectively the following equations of motion: 
\begin{equation}
\begin{array}{l}
\dot{\varphi}^a=\omega^{ab}\partial_bH \medskip \\
\dot{c}^a=\omega^{ad}\partial_d\partial_bH c^b \medskip \\
\dot{\bar{c}}_b=-\bar{c}_a\omega^{ad}\partial_d\partial_bH \medskip \\
\dot{\lambda}_b=-\omega^{ad}\partial_d\partial_bH\lambda_a-i\bar{c}_a\omega^{ad}
\partial_d\partial_f\partial_bHc^f. \label{20-1}
\end{array}
\end{equation}
They could be obtained also via the $\widetilde{\cal H}$ of (\ref{14-1b}) by postulating the following {\it e}xtended {\it P}oisson {\it b}rackets (epb):
\bea
\displaystyle \{\varphi^a,\lambda_b\}_{epb}=\delta_b^a, && \qquad  \{\bar{c}_b,c^a\}_{epb}=-i\delta_b^a, \medskip \label{3.2} \\
&&\displaystyle  \hspace{-1cm} \{\varphi^a,\varphi^b\}_{epb}=0. \label{3.3}
\label{20-2}
\eea 
Note that Eq. (\ref{3.3}) is different from the {\it P}oisson {\it b}rackets (pb)
defined in the {\it non-extended} phase space which are:
\be
\displaystyle \{ \varphi^a,\varphi^b\}_{pb}=\omega^{ab}. \label{3-3bis}
\ee
So we are basically working in an extended phase space. Its precise geometrical structure is the one of a double bundle over the basic phase space ${\cal M}$ coordinatized by the variables $\varphi^a$. This
double bundle has been studied in detail in Ref. \cite{quattro}.

The reader may dislike the pletora of variables $(\varphi^a,\lambda_a,c^a,\bar{c}_a)$
that make up the {\it extended phase space} described above. Actually, thanks to the beautiful geometry 
underlying this space, we can assemble together the $8n$ variables $(\varphi^a,\lambda_a,c^a,\bar{c}_a)$
in a single object as if they were the components of a multiplet. In order to do that we have first to
introduce two Grassmann partners $(\theta,\bar{\theta})$ of the time $t$. The triplet
\be
\displaystyle (t,\theta,\bar{\theta}) \label{23-0}
\ee
is known as {\it supertime} and it is, for the point particle dynamics, the analog of the {\it superspace} 
introduced in supersymmetric field theories \cite{nove}. The object that we mentioned above and which
assembles together the $8n$ variables $(\varphi,\lambda,c, \bar{c})$ is defined as
\be
\displaystyle \Phi^a(t,\theta,\bar{\theta})\equiv \varphi^a(t)+\theta c^a(t)+\bar{\theta}
\omega^{ab}\bar{c}_b(t)+i\bar{\theta}\theta\omega^{ab}\lambda_b(t). \label{21-1}
\ee
We could call the $\Phi^a$ {\it superphase space} variables because their first components $\varphi^a$ 
are the standard phase space variables of the system.
The Grassmann variables $\theta$, $\bar{\theta}$ are complex in the sense in which the operation of complex conjugation can be defined \cite{dieci} for Grassmann variables. Further details can be found in Appendix A, where we also study the dimensions of these Grassmann variables. The main result is that, even if there is a lot of freedom in choosing these dimensions, the combination $\theta\bar{\theta}$ has always the dimensions of an action. Using the superfields the relations (\ref{3.2})-(\ref{3.3}) can be written in a compact form as:
\begin{displaymath}
\{\Phi^a(t,\theta, \bar{\theta}),\Phi^b(t,\theta^{\prime},\bar{\theta}^{\prime})\}_{epb}=
-i \omega^{ab}\delta(\bar{\theta}-\bar{\theta}^{\prime})\delta(\theta-\theta^{\prime}),
\end{displaymath}
which is the extended phase space analog of (\ref{3-3bis}). We would like to draw the attention of the reader to the fact that our $epb$ among superfields are different than those postulated in Ref. \cite{batalin}. 

Despite the formal unification of the $8n$ variables $(\varphi,c,\bar{c},\lambda)$
into the $2n$ objects $\Phi^a(t,\theta,\bar{\theta})$ of (\ref{21-1}), the reader may still wonder
why we need $8n$ variables when we know that CM can be described using only the basic phase space variables $\varphi^a$. The answer lies in the fact that we want an object, like $\widetilde{\cal H}$, which makes at the same time the evolution of the points of the phase space $\varphi$ and of the forms $d\varphi$. But we know that the evolution of the forms $d\varphi$ can actually be derived from the evolution of the points by doing its first variation and this implies that somehow the variables $(\varphi,c,\bar{c},\lambda)$ are redundant. This redundancy is signalled by a set of universal symmetries present in our formalism \cite{due}\cite{quattro}. The charges associated to these symmetries are listed below in their operatorial form:
\be
\begin{array}{l}
\displaystyle \hat{Q}_{\scriptscriptstyle BRS}\equiv i\hat{c}^a\hat{\lambda}_a, \medskip \\
\displaystyle \hat{\bar{Q}}_{\scriptscriptstyle BRS}\equiv i\hat{\bar{c}}_a
\omega^{ab}\hat{\lambda}_b, \medskip \\
\displaystyle \hat{Q}_{\scriptscriptstyle H} \equiv \hat{Q}_{\scriptscriptstyle BRS}-
\hat{c}^a\partial_aH, \medskip\\
\displaystyle \hat{\bar{Q}}_{\scriptscriptstyle H} \equiv \hat{\bar{Q}}_{\scriptscriptstyle BRS}+
\hat{\bar{c}}_a\omega^{ab}\partial_bH, \medskip\\
\displaystyle \hat{K} \equiv \frac{1}{2}\omega_{ab}\hat{c}^a\hat{c}^b, \medskip\\
\displaystyle \hat{\bar{K}} \equiv \frac{1}{2}\omega^{ab}\hat{\bar{c}}_a\hat{\bar{c}}_b, \medskip\\
\displaystyle \hat{Q}_g \equiv \hat{c}^a\hat{\bar{c}}_a. \label{23-1}
\end{array}
\ee 
We have put on these charges some labels (like $BRS$ for Becchi-Rouet-Stora or ``$g$" for ghost) because they resemble some charges that carry the same name and are present in gauge theories. The two charges we are most interested in are $\hat{Q}_{\scriptscriptstyle H}$ and 
$\hat{\bar{Q}}_{\scriptscriptstyle H}$ because they make up a ``{\it universal}"
$N=2$ supersymmetry (susy), present in any system. In fact their anticommutator is:
\begin{displaymath}
\displaystyle [\hat{Q}_{\scriptscriptstyle H},\hat{\bar{Q}}_{\scriptscriptstyle H}]=
2i\hat{\widetilde{\cal H}}. \label{23-2}
\end{displaymath}
The charges in (\ref{23-1}) act via graded commutators on the variables $(\hat{\varphi}^a,\hat{c}^a,
\hat{\bar{c}}_a,\hat{\lambda}_a)$ which can be considered as the ``{\it target space}" variables (in modern language), while the ``{\it base space}" is given by the supertime $(t,\theta,\bar{\theta})$. It is then natural to ask whether the operators (\ref{23-1}) are the ``representation" on the target space of some differential operator acting on the base space. The answer is yes and the form of these differential
operators is easy to obtain if we impose on the superfield (\ref{21-1}) to be a {\it scalar}
under the symmetry (\ref{23-1}), i.e.\footnote{We have put an hat on the $\Phi^a$ defined in (\ref{21-1}) to indicate that we have turned into operators the variables $\varphi^a,c^a,\bar{c}_a,\lambda_a$ appearing in (\ref{21-1}), using the rules derived in Sec. 2.}:
\be
\displaystyle \hat{\Phi}^{\prime a}(t^{\prime},\theta^{\prime},\bar{\theta}^{\prime})=
\hat{\Phi}^a(t,\theta,\bar{\theta}). \label{24-1}
\ee
If we indicate with $\hat{O}$ any of the charges in (\ref{23-1}) and with $\hat{\cal O}$ the corresponding differential operator on the base space, then relation (\ref{24-1}) is equivalent to the following one
\begin{displaymath}
\displaystyle \hat{\cal O}\Phi^a=[\hat{\Phi}^a,\hat{O}]. \label{24-2}
\end{displaymath}
From this equation it is easy to derive the form of the various operators $\hat{\cal O}$. We indicate below the expressions of some of them:
\be
\begin{array}{l}
\displaystyle \hat{\cal Q}_{\scriptscriptstyle BRS}=-\frac{\partial}{\partial \theta} \medskip\\
\displaystyle \hat{\bar{{\cal Q}}}_{\scriptscriptstyle BRS}=\frac{\partial}{\partial \bar{\theta}} \medskip\\
\displaystyle \hat{\cal Q}_{\scriptscriptstyle H}=
-\frac{\partial}{\partial \theta}-\bar{\theta}\frac{\partial}{\partial t} \medskip\\
\displaystyle \hat{\bar{{\cal Q}}}_{\scriptscriptstyle H}=\frac{\partial}{\partial \bar{\theta}}+
\theta \frac{\partial}{\partial t} \medskip\\
\displaystyle \hat{\widetilde{\cal H}}=i\frac{\partial}{\partial t}. \label{25-1}
\end{array}
\ee
From these expressions one sees that the BRS and anti-BRS operators act as translation operators in $\theta$ and $\bar{\theta}$. Instead the supersymmetry operators $\hat{\cal Q}_{\scriptscriptstyle H}$ and $\hat{\bar{{\cal Q}}}_{\scriptscriptstyle H}$ turn $t$ into combinations of $t$ with $\theta$ and $\bar{\theta}$, that is why we called $\theta$ and $\bar{\theta}$ ``partners" of $t$. Once they are applied in sequence, these susy operators perform a time translation, as shown by the commutator
\begin{displaymath}
\displaystyle [\hat{\cal Q}_{\ssc H},\hat{\bar{{\cal Q}}}_{\ssc H}]=-2\frac{\partial}{\partial t}.
\end{displaymath} 
The supertime (\ref{23-0}) is somehow made of three ``coordinates"
$(t,\theta,\bar{\theta})$. So two different instants in supertime will have respectively coordinates
$(t_{\ssc 1},\theta_{\ssc 1},\bar{\theta}_{\ssc 1})$ and $(t_{\ssc 2},\theta_{\ssc 2},\bar{\theta}_{\ssc 2})$. A natural question 
to ask is the following: is there an interval between those two {\it super-instants} which is 
left invariant by the supersymmetry transformations? The answer is yes and actually there is more than one such interval. For example we can take:
\be
\displaystyle S\equiv t_{\ssc 2}-t_{\ssc 1}+\theta_{\ssc 2}\bar{\theta}_{\ssc 1}-\theta_{\ssc 1}\bar{\theta}_{\ssc 2}. \label{26-1}
\ee
Eq. (\ref{26-1}) generalizes the usual 
interval of time $(t_{\ssc 2}-t_{\ssc 1})$, which is invariant under a global time translation.
Also the interval (\ref{26-1}) is invariant under global time translations just because, being invariant
under susy, it is also invariant if susy is performed twice, which is a time translation.
In Appendix B we will specify better what it means that (\ref{26-1}) is invariant under supersymmetry
and we will present other intervals of supertime. 
All this formalism on time and supertime is familiar to people working on supersymmetry but it may not be so familiar to those working on issues related with quantization to which this paper is addressed. 

A further aspect of the Grassmann partners of time $(\theta,\bar{\theta})$
which is worth being explored is the following one. We know that, for an operator theory, we can define either the Heisenberg picture or the Schr\"odinger one. In the KvN version of CM the operators, which will be indicated respectively as $\hat{O}_{\ssc H}(t)$ and $\hat{O}_{\ssc S}$ in the two pictures, are related to each other in the following manner:
\begin{displaymath}
\displaystyle \hat{O}_{\ssc H}(t)\equiv \exp \Bigl[i\hat{\widetilde{\cal H}}t\Bigr] 
\hat{O}_{\ssc S} \exp \Bigl[-i\hat{\widetilde{\cal H}}t\Bigr].
\label{27-1}
\end{displaymath}
One question to ask is what happens if we build the Heisenberg picture associated to the partners of time
$\theta,\bar{\theta}$. The analog of the time translation operator $\hat{\widetilde{\cal H}}$ for $\theta$ and $\bar{\theta}$ is given by $\hat{Q}_{\scriptscriptstyle BRS}$ and 
$\hat{\bar{Q}}_{\scriptscriptstyle BRS}$ respectively (see Eqs. (\ref{23-1})-(\ref{25-1})) and so the Heisenberg picture in $\theta$, $\bar{\theta}$ of an operator $\hat{O}_{\ssc S}$ is:\footnote{With the scalar product under which the Grassmann operators $\hat{c}$ and $\hat{\bar{c}}$ are Hermitian \cite{sette}, the operators $\exp \hat{\bar{Q}} \bar{\theta}$ and $\exp \theta \hat{Q}$ are unitary, see Appendix A for further details.}
\be
\displaystyle \hat{O}_{\ssc H}(\theta,\bar{\theta}) \equiv \exp \left[\theta \hat{Q}+\hat{\bar{Q}}\bar{\theta}\right]
\hat{O}_{\ssc S} \exp \left[-\theta \hat{Q}-\hat{\bar{Q}}\bar{\theta}\right], \label{connect}
\ee
where we have dropped the ``BRS" suffix from $\hat{Q}$ and $\hat{\bar{Q}}$. A simple example to start from is the phase space operator $\hat{\varphi^a}(t)$ which does not depend on $\theta,\bar{\theta}$, so it could be considered as an operator in the Schr\"odinger picture with respect to $\theta,\bar{\theta}$ and in the Heisenberg picture with respect to $t$. Its Heisenberg picture version in $\theta,\bar{\theta}$
can be worked out easily (see Appendix C) and the result is:
\begin{displaymath}
\hat{\varphi}^a_{\ssc H}(t)\equiv \exp \Bigl[ \theta\hat{Q}+\hat{\bar{Q}}\bar{\theta}\Bigr] \hat{\varphi}_{\ssc S}^a(t)
\exp \Bigl[ -\theta \hat{Q}-\hat{\bar{Q}}\bar{\theta}\Bigr]
=\hat{\Phi}^a(t,\theta,\bar{\theta}).
\end{displaymath}
This means the {\it superphase space operators} $\hat{\Phi}^a$ {\it can be considered as the Heisenberg picture} version in $\theta$, $\bar{\theta}$ of the phase space operators.
The same holds for any function $G$ of the operators $\hat{\varphi}$, i.e.:
\be
\displaystyle \exp \Bigl[\theta \hat{Q}+\hat{\bar{Q}}\bar{\theta}\Bigr] G(\hat{\varphi}^a)
\exp \Bigl[-\theta\hat{Q}-\hat{\bar{Q}}\bar{\theta}\Bigr] =G(\hat{\Phi}^a), \label{28-1}
\ee
see Appendix C for details.
In particular, if the function $G(\hat{\varphi}^a)$ is the Hamiltonian $H(\hat{\varphi}^a)$ we get from (\ref{28-1})
\be
\displaystyle SH(\hat{\varphi}^a)S^{-1} =H[\hat{\Phi}^a] \label{29-1}
\ee
where $S=\exp [\theta \hat{Q} +\hat{\bar{Q}}\bar{\theta}]$. 

At this point it is instructive to expand the RHS of (\ref{29-1}) in terms of $\theta$, $\bar{\theta}$. We get:
\be
\displaystyle H[\hat{\Phi}^a]=H[\hat{\varphi}^a]+\theta \hat{N}+\hat{\bar{N}}\bar{\theta}-
i\bar{\theta}\theta \hat{\widetilde{\cal H}} \label{29-2}
\ee
where 
\begin{displaymath}
\hat{N}=\hat{c}^a\partial_aH(\hat{\varphi}), \qquad \hat{\bar{N}}=
\hat{\bar{c}}_a\omega^{ab}\partial_bH(\hat{\varphi})
\end{displaymath}
are further conserved charges \cite{quattro}. The expansion (\ref{29-2}) holds also if we replace 
the operators with the corresponding $c$-number variables, i.e.:
\be
\displaystyle H[\Phi^a]=H[\varphi^a]+\theta N+\bar{N}\bar{\theta}-i\bar{\theta}\theta
\widetilde{\cal H}(\varphi^a,\lambda_a,c^a,\bar{c}_a). \label{29-3}
\ee
It is interesting to note that the first term in the expansion in $\theta,\bar{\theta}$
on the RHS of (\ref{29-3}) is $H(\varphi^a)$ which generates the dynamics in the standard 
{\it phase space} $\varphi^a$ while the last term is $\widetilde{\cal H}$, which generates 
the dynamics in the extended phase space $(\varphi^a,\lambda_a,c^a,\bar{c}_a)$.

The last thing we want to do in this section is to study the equations of motion in the Heisenberg 
picture with respect to $\theta$, $\bar{\theta}$. Using the commutators (\ref{14-2}), the
equations of motion for ${\varphi}^a$ are
\begin{displaymath}
\dot{\varphi}^a=i[\hat{\widetilde{\cal H}},\varphi^a]. \label{30-1} 
\end{displaymath}
Passing now to the Heisenberg picture in $\theta$ and $\bar{\theta}$, we get
\begin{displaymath}
\displaystyle 
S\dot{\varphi}^aS^{-1}=iS[\hat{\widetilde{\cal H}},\varphi^a]S^{-1} \;\; \Longrightarrow \;\;
\dot{\Phi}^a=i[S\hat{\widetilde{\cal H}}S^{-1},\Phi^a] \;\; \Longrightarrow \;\; \dot{\Phi}^a=i[\hat{\widetilde{\cal H}},\Phi^a]. \label{30-2}
\end{displaymath}
All the steps above are trivial except the last one, i.e. 
$S\hat{\widetilde{\cal H}}S^{-1}=\hat{\widetilde{\cal H}}$, which is due to the fact that both $\hat{Q}$
and $\hat{\bar{Q}}$ entering $S$ commute with $\hat{\widetilde{\cal H}}$. The beauty of the equation of motion 
\be
\displaystyle \dot{\Phi}^a=i[\hat{\widetilde{\cal H}},\Phi^a] \label{31-1}
\ee
is that it encapsulates in a single equation all the equations of motion (\ref{20-1}) 
for the $8n$ variables $(\varphi^a,c^a,\bar{c}_a,\lambda_a)$, as can be proved by expanding 
(\ref{31-1}) in $\theta$, $\bar{\theta}$. We could call (\ref{31-1}) 
the super-Heisenberg equation of motion.

We want to conclude this section by noting that our equations of motion (\ref{31-1}) are
different than those developed in Ref. \cite{batalin}. Ours are the same 
as those which were derived in Ref. \cite{due} and in (\ref{20-1}) by a variational principle from the Lagrangian $\widetilde{\cal L}$. 

\section{Dequantization in the $q$ and $p$-polarizations and supertime}

In this section, which is the central one for our project, we will study 
the role of the superphase space variables $\Phi^a$ at the Lagrangian and path 
integral level.

We have seen that, for what concerns the Hamiltonians, relation (\ref{29-3}) holds:
\begin{displaymath}
H[\Phi]=H[\varphi]+\theta N+\bar{N}\bar{\theta}-i\bar{\theta}\theta \widetilde{\cal H}
\end{displaymath}
which implies that
\be
\displaystyle i \int d\theta d\bar{\theta} H[\Phi]=\widetilde{\cal H}. \label{32-1}
\ee
An analog of this relation at the Lagrangian level does not hold exactly. The reason, as explained in detail in Appendix C, is the presence in the Lagrangian of the kinetic terms
$p\dot{q}$, which are not present in $H$, i.e.:
\begin{displaymath}
\displaystyle L(p,q)=p\dot{q}-H(p,q). \label{32-2}
\end{displaymath}
Replacing $q$ and $p$ in the Lagrangian $L$ with the superphase space variables $\Phi^q$ and $\Phi^p$, the analog of Eq. (\ref{32-1}) becomes the following:
\be
\displaystyle i\int d\theta d\bar{\theta} L(\Phi)=\widetilde{\cal L}-
\frac{d}{dt}(\lambda_{p_i}p_i+i\bar{c}_{p_i}c^{p_i}), \label{33-1}
\ee
where $\widetilde{\cal L}$ is the Lagrangian of the CPI given by Eq. (\ref{14-1a}),
and the $\lambda_{p_i}$, $\bar{c}_{p_i}$ and $c^{p_i}$ are the second half of the variables
$\lambda_a$, $\bar{c}_a$ and $c^a$. From now on we will change our notation for the superphase space variables: instead of writing 
\begin{displaymath}
\displaystyle \Phi^a\equiv \varphi^a+\theta c^a+\bar{\theta}\omega^{ab}\bar{c}_b+i\bar{\theta}\theta
\omega^{ab}\lambda_b
\end{displaymath}
we will explicitly indicate the $q$ and $p$ components in the following manner:
\be
\displaystyle \Phi^a=\begin{pmatrix} Q_i \cr P_i
\end{pmatrix}\equiv \begin{pmatrix} q_i \cr p_i \end{pmatrix}+\theta \begin{pmatrix}c^{q_i} \cr
c^{p_i} \end{pmatrix}+\bar{\theta}\begin{pmatrix} \bar{c}_{p_i} \cr -\bar{c}_{q_i}
\end{pmatrix}+i\bar{\theta}\theta \begin{pmatrix} \lambda_{p_i} \cr
-\lambda_{q_i}\end{pmatrix}, \label{33-2}
\ee
where $i=(1,\cdots, n)$, and $a=(1,\cdots, 2n)$. 

Going now back to (\ref{33-1}) we could write it as follows: 
\be
\displaystyle \widetilde{\cal L}=i\int d\theta d\bar{\theta} L[\Phi]+\frac{d}{dt}(\lambda_pp
+i\bar{c}_pc^p), \label{33-3}
\ee 
where we have dropped the indices ``$i$" appearing on the extended phase space variables. The expression of $\widetilde{\cal L}$
which appears in (\ref{33-3}) can be used in (\ref{19-3}) and we get
\bea
\displaystyle \langle \varphi,c,t|\varphi_{\ssc 0},c_{\ssc 0},t_{\ssc 0}\rangle &=& 
\int {\mathscr D}^{\prime\prime}\varphi{\mathscr D}\lambda{\mathscr D}^{\prime\prime}c{\mathscr D}
\bar{c} \, \exp i\int_{t_0}^t d\tau \widetilde{\cal L}= \nonumber\\
&=& \int {\mathscr D}^{\prime\prime}\varphi {\mathscr D}\lambda {\mathscr D}^{\prime\prime}c
{\mathscr D}\bar{c} \, \exp \Bigl[ i\int_{t_0}^t i d\tau d\theta d\bar{\theta}L[\Phi]+(\textrm{s.t.})\Bigr]
\label{34-1}
\eea
where (s.t.) indicates the surface terms, which come from the total derivative appearing on the RHS of (\ref{33-3}) and has the form
\be
\displaystyle (\textrm{s.t.})=i\lambda_pp-i\lambda_{p_{\ssc 0}}p_{\ssc 0}-
\bar{c}_pc^p+\bar{c}_{p_{\ssc 0}}c^{p_{\ssc 0}}. \label{34-2}
\ee 
We indicate with $p_{\ssc 0}$ the $n$ components of the initial momenta.
The surface terms present in (\ref{34-1}) somehow spoil the beauty of formula (\ref{34-1}) but we can get rid of them by changing the basis of our Hilbert space.
We showed in Sec. 2 that, besides the basis $| \varphi,c\rangle=| q,p,c^q,c^p\rangle$, we could
introduce the ``mixed" basis defined in (\ref{15-2}) by the states: $|q,\lambda_p,
c^q,\bar{c}_p\rangle$. We can then pass from the transition amplitude
$\langle q,p,c^q,c^p,t|q_{\ssc 0},p_{\ssc 0},c^{q_{\ssc 0}},\bar{c}_{p_{\ssc 0}},t_{\ssc 0}\rangle$ of (\ref{34-1}) 
to the mixed one $\langle q,\lambda_p,c^q,\bar{c}_p,t|q_{\ssc 0},\lambda_{p_{\ssc 0}},
c^{q_{\ssc 0}},\bar{c}_{p_{\ssc 0}}, t_{\ssc 0}\rangle$, which 
are related to each other as follows:
\bea
&& \displaystyle
\langle q,\lambda_p,c^q,\bar{c}_p,t|q_{\ssc 0},\lambda_{p_{\ssc 0}},c^{q_{\ssc 0}}, 
\bar{c}_{p_{\ssc 0}},t_{\ssc 0}\rangle= \nonumber\\
&& \displaystyle =\int dp\,dp_{\ssc 0}\,dc^p\,dc^{p_{\ssc 0}}\, e^{-i\lambda_pp}e^{\bar{c}_pc^p}
\langle q,p,c^q,c^p,t|q_{\ssc 0},p_{\ssc 0},c^{q_{\ssc 0}},c^{p_{\ssc 0}},t_{\ssc 0}\rangle 
e^{i\lambda_{p_{\ssc 0}}p_{\ssc 0}}
e^{-\bar{c}_{p_{\ssc 0}}c^{p_{\ssc 0}}}. \label{35-1}
\eea 
Replacing on the RHS of this formula the kernel $\langle q,p,c^q,c^p,t|q_{\ssc 0},p_{\ssc 0}, c^{q_{\ssc 0}},c^{p_{\ssc 0}},t_{\ssc 0}\rangle$ with its path integral expression given in (\ref{34-1}), we get the very {\it neat} expression:
\be
\displaystyle \langle q,\lambda_p,c^q,\bar{c}_p,t|q_{\ssc 0},\lambda_{p_{\ssc 0}}, 
c^{q_{\ssc 0}},\bar{c}_{p_{\ssc 0}},t_{\ssc 0}\rangle=
\int {\mathscr D}^{\prime\prime}Q{\mathscr D}P\, \exp \Bigl[i\int_{t_0}^t id\tau d\theta d\bar{\theta}L(\Phi)\Bigr]
\label{35-2}
\ee
where 
\be
\displaystyle {\mathscr D}^{\prime\prime}Q{\mathscr D}P\equiv {\mathscr D}^{\prime\prime}
q{\mathscr D}p{\mathscr D}^{\prime\prime}\lambda_{p}{\mathscr D}\lambda_{q}
{\mathscr D}^{\prime\prime}c^{q}{\mathscr D}c^{p}{\mathscr D}^{\prime\prime}\bar{c}_{p}
{\mathscr D}\bar{c}_{q}. \label{35-3} 
\ee
We would like to point out three ``interesting" aspects of Eq. (\ref{35-2}). 
\begin{enumerate}
\item[{\bf 1.}] Note that the surface terms of Eq. (\ref{34-1}) have disappeared in Eq. (\ref{35-2});
\item[{\bf 2.}] The measure in the path integral (\ref{35-2}) is the same
as the measure of the quantum path integral \cite{uno}, which is
\be
\displaystyle \langle q,t|q_{\ssc 0},t_{\ssc 0}\rangle =\int {\mathscr D}^{\prime\prime}q{\mathscr D}p \;
\exp \frac{i}{\hbar} \int d\tau\, L[\varphi]. \label{36-1} 
\ee
What we mean is that both in (\ref{35-2}) and in (\ref{36-1}) the integration in $p$ and $P$ is done
even over the initial and final variables while the integration in $q$ and $Q$ is done only over the intermediate 
points between the initial and final ones. The reason for the notation (\ref{35-3}) should be clear from the 
fact that the superphase space variables $Q$, $P$ are defined as
\be
\begin{array}{l}
Q \equiv q+\theta c^{q} +\bar{\theta}\bar{c}_{p}+i\bar{\theta}\theta \lambda_{p} \medskip \\
P\equiv p+\theta c^{p}-\bar{\theta}\bar{c}_{q}-i\bar{\theta}\theta \lambda_{q},
\label{36-2}
\end{array}
\ee
so the integration over $Q$, $P$ means the integration over the elements $(q,c^{q},
\bar{c}_{p},\lambda_{p})$ and $(p,c^{p},
\bar{c}_{q},\lambda_{q})$ which make up $Q$ and $P$ respectively;
\item[{\bf 3.}] The function $L$, which enters both the QM path integral (\ref{36-1}) 
and the CM one (\ref{35-2}), is the same. The only difference is that in QM (\ref{36-1}) the variables entering $L$ are the normal phase space variables while in CM (\ref{35-2}) they are the superphase space variables $\Phi^a=(Q,P)$.
\end{enumerate}
Let us now proceed by noticing that formally we can rewrite (\ref{35-2}) as
\be
\displaystyle \langle Q,t|Q_{\ssc 0},t_{\ssc 0}\rangle \equiv \int {\mathscr D}^{\prime\prime}Q{\mathscr D}P\,
\exp \left[i\int_{t_0}^t i d\tau d\theta d\bar{\theta} L[\Phi]\right] \label{36-3}
\ee
where we have defined the ket $|Q\rangle$ as the common eigenstate of the operators
$\hat{q}$, $\hat{\lambda}_p$, $\hat{c}^q$, $\hat{\bar{c}}_p$:
\be
\begin{array}{l}
\hat{q}|Q\rangle =q|Q\rangle,  
\medskip \\
\hat{c}^q|Q\rangle =c^q|Q\rangle,  
\end{array}
\qquad \quad
\begin{array}{l}
\hat{\lambda}_p|Q\rangle =\lambda_p|Q\rangle, 
\medskip \\
\hat{\bar{c}}_p|Q\rangle =\bar{c}_p|Q\rangle. \label{37-1}
\end{array}
\ee
So $|Q\rangle$ can be ``identified" with the state $|q,\lambda_p,c^q,\bar{c}_p\rangle$ which appears
in (\ref{35-2}). The reader may not like this notation because in (\ref{36-2})  $Q$ contains the Grassmann variables $\theta$ 
and $\bar{\theta}$ which do not appear at all in (\ref{37-1}). Actually, from (\ref{37-1}) we can derive that the state $|Q\rangle$ is also an eigenstate of the supervariable $\hat{Q}$ obtained by turning the expression (\ref{36-2}) into an operator, i.e.:
\be
\displaystyle \hat{Q}(\theta,\bar{\theta})|Q\rangle =Q(\theta,\bar{\theta})|Q\rangle. \label{37-2}
\ee
This relation is just a simple consequence of (\ref{37-1}) as can be proved by expanding in $\theta$ and $\bar{\theta}$ both $\hat{Q}$ and $Q$ in (\ref{37-2}). One immediately sees that 
the variables $\theta$ and $\bar{\theta}$ make their appearance not in the state $|Q\rangle$ but in its 
eigenvalue $Q$ and in the operator $\hat{Q}$. We are now ready to compare (\ref{36-1}) and (\ref{36-3}). One is basically the central element of QM:
\be
\displaystyle \langle q,t|q_{\ssc 0}, t_{\ssc 0}\rangle =\int {\mathscr D}^{\prime\prime} q{\mathscr D}p
\exp \left[\frac{i}{\hbar} \int_{t_0}^t d\tau L[\varphi]\right], \label{38-1}
\ee 
while the other is the central element of CM (formulated \`a la KvN or \`a la CPI):
\be
\displaystyle \langle Q,t|Q_{\ssc 0},t_{\ssc 0}\rangle =\int {\mathscr D}^{\prime\prime}Q{\mathscr D}P
\exp \left[ i\int_{t_0}^t i d\tau d\theta d\bar{\theta} \,L[\Phi]\right]. \label{38-2}
\ee
By just looking at (\ref{38-1}) and (\ref{38-2}), it is now easy to give some simple {\it rules}, which turn the {\it quantum} transition amplitude (\ref{38-1}) into the {\it classical} one (\ref{38-2}). The rules are:
\begin{enumerate}
\item[{\bf 1)}] {\it Replace in the QM case the phase space variables} $(q,p)$ {\it everywhere with the
superphase space ones} $(Q,P)$;
\item[{\bf 2)}] {\it Extend the time integration to the supertime integration multiplied by} $\hbar$
\be
\displaystyle \int d\tau \, \longrightarrow \, i\hbar \int d\tau d\theta d\bar{\theta}. \label{39-1}
\ee
\end{enumerate}
The reason for the appearance of the ``$i$" on the RHS of (\ref{39-1}) is related to the complex nature of the Grassmann variables $\theta$ and $\bar{\theta}$, as explained in Appendix A. The reason for the appearance of $\hbar$ instead is related to the fact that in (\ref{38-2}), which is CM, there is no $\hbar$ and so in (\ref{39-1}) we need to introduce an $\hbar$ in order to cancel the one of Eq. (\ref{38-1}). From the dimensional point of view formula (\ref{39-1}) 
is correct because, as shown in Appendix A, the dimensions of $d\theta d\bar{\theta}$ are just the inverse of an action canceling in this manner the dimension of $\hbar$ appearing in front of the RHS in (\ref{39-1}).
This implies that both the LHS and the RHS of (\ref{39-1}) have the dimensions of a time. We will call the rules {\bf 1)} and {\bf 2)} above as dequantization rules.

Note that these {\it dequantization} rules are {\it not} the semiclassical or WKB limit of QM.
In fact we are not sending $\hbar \to 0$ in (\ref{38-1}) and what we get is not QM in the 
leading order in $\hbar$, like in the WKB method, but exactly CM in the KvN or CPI formulation.
We named this procedure ``dequantization" because it is the inverse of ``quantization" in the sense
that, while quantization is a set of rules which turn CM into QM, our rules {\bf 1)} and {\bf 2)}
turn QM into CM.
We called `` {\it geometrical}" our procedure because it basically consists of a geometrical 
{\it extension} of both the base space given by time $t$, into the supertime $(t,\theta,\bar{\theta})$ 
and of the target space, which is phase space $(q,p)$ in QM, into a superphase space $(Q,P)$ in CM.
We used the expressions ``{\it base space}" and ``{\it target space}", as it is done nowadays
in strings and higher dimensional theories, where procedures of dimensional extension or dimensional
contraction are very often encountered. In those theories the procedures of dimensional extension 
is introduced to give a geometrical basis to the many extra fields present in grand-unified theories,
while the procedure of dimensional contraction is needed to come back to our four-dimensional world.
We find it amazing and thought-provoking that even the procedure of quantization (or dequantization)
can be achieved via a dimensional contraction (or extension).

In previous papers \cite{cinque} we have given brief presentations of these ideas but there we explored the inverse route, that is the one of {\it quantization}, which is basically how to pass from (\ref{38-2}) to (\ref{38-1}). This goal is achieved by a sort of dimensional reduction from the supertime $(t,\theta,\bar{\theta})$ to the time $t$, and from the superphase space $(Q,P)$ to the phase space $(q,p)$. In those papers \cite{cinque}
we thought of implementing the supertime reduction by inserting a 
$\delta(\bar{\theta})\delta(\theta)/\hbar$ into the weight appearing in (\ref{38-2}) but we found this method
a little bit awkward and that is why in this paper we have preferred to explore the opposite 
route that is the one of {\it dequantization} which is brought about by a dimensional extension. Even if awkward to implement, the quantization route from (\ref{38-2}) to (\ref{38-1}) can be
compared with a well-known method of quantization known in the literature as {\it geometric quantization}
(GQ) \cite{sei}. We will not review it here but suffice it to say that it starts from the so-called  ``prequantization space" (which is our space of KvN states $\psi(q,p)$), and from the {\it Lie derivative
of the Hamiltonian flow} (which is our $\hat{\widetilde{\cal H}}$ of equation (\ref{16-3})), and, through a long set of steps, it builds up the Schr\"odinger operator and the Hilbert space of QM. Basically, in going from (\ref{38-2}) to (\ref{38-1}), we do the same because we go from the weight
$\exp i \int i dt d\theta d\bar{\theta} L[\Phi]$, which is the evolution via the Lie derivative 
operator $\exp -i\hat{\widetilde{\cal H}} t$, to the weight $\exp \frac{i}{\hbar}\int
L[\varphi]$, which reproduces the evolution via the Schr\"odinger operator 
$ \exp -i\frac{\hat{H}}{\hbar}t$. Regarding the states we go from the KvN states 
$|Q\rangle $ to the Schr\"odinger ones $| q\rangle$ by just sending $\theta,\bar{\theta} \to 0$ in Eq. (\ref{37-2}). The difference with respect to GQ is that our KvN states contain
also the Grassmann variables $c$ and $\bar{c}$, which are not contained in the prequantization states of GQ. In GQ the reduction of the KvN states to the quantum ones is achieved via a procedure called
``polarization" while the transformation of the Lie derivative into the quantum Schr\"odinger
operator is achieved via a totally different procedure, see Ref. \cite{sei} for details. In our opinion it is somehow unpleasant that in GQ states and operators are ``quantized" via two totally different procedures. This is not so anymore in our functional approach, which brings (\ref{38-2}) into (\ref{38-1}). It is in fact the dimensional reduction both in $t$ 
\be
\displaystyle i\hbar \int dt d\theta d\bar{\theta} \, \longrightarrow \, \int dt \label{43-1}
\ee 
and in phase space
\be
\displaystyle (Q,P) \, \longrightarrow \, (q,p), \label{43-2}
\ee
which at the same time produces the right operators (from the classical Lie derivative to the Schr\"odinger operator)
and the right states (from the KvN states $|Q\rangle$ to the quantum ones $| q\rangle$). So we do not need two
different procedures for operators and states but just a single one. Actually the dimensional reduction contained in (\ref{43-1}) and (\ref{43-2}) can be combined in a single operation, that is the one of {\it shrinking to zero} the variables $\theta,\bar{\theta}$, i.e.: $(\theta,\bar{\theta})\,\to \, 0$.
This not only brings the integration $\int dt d\theta d\bar{\theta}$ to $\int dt$ but, remembering 
the form of $(Q,P)$ i.e. (\ref{36-2}), it also brings 
\begin{displaymath}
\displaystyle Q\, \longrightarrow \, q.
\end{displaymath} 
Because of this, it reduces the KvN states $| Q\rangle$ to the quantum ones $| q\rangle$, which are a basis for the quantum Hilbert space in the Schr\"odinger representation. 
Note the difference with the GQ procedure: there one starts with the states $| q,p\rangle$ and the ``$p$" is removed through a long set of steps, known as polarization \cite{sei}. In our approach instead we first replace the ``$p$" in the $| q,p \rangle$ states with the $\lambda_p$ via the {\it Fourier transform} presented in (\ref{35-1}) and then remove the $\lambda_p$ by {\it sending} $(\theta,\bar{\theta})\to 0$. We want to stress again that the same two steps, {\bf 1)} {\it Fourier transform} and {\bf 2)} {\it sending} $\theta,\bar{\theta} \to 0$, which polarizes the states, are the same ones which turn the {\it classical } evolution into the {\it quantum} one. In fact step {\bf 1)} takes away the surface terms in (\ref{34-1}) bringing the weight to be of the same form as the quantum one and step {\bf 2)}, sending $(\theta,\bar{\theta})\to 0$, brings the classical weight into the quantum one. We feel that this coincidence of the two procedures, for the states and the operators, was not noticed in GQ because there they did not use the partners of time $(\theta,\bar{\theta})$ and the functional approach. This coincidence is quite interesting because (besides a trivial Fourier transform) it boils down to be a {\it geometrical} operation: the dimensional reduction from supertime $(t,\theta,\bar{\theta})$ to time $t$. We feel that this is really the {\it geometry} at the heart of geometric quantization and of quantum mechanics in general. 

The reader anyhow may suspect that the simple operation of sending $\theta,\bar{\theta} \, \to \, 0$ may well produce both the right quantum evolution and at the same time polarize the states but that this coincidence takes place only in the Schr\"odinger polarization. We will prove that this is not so. Below we shall show that all this procedure works also in the momentum polarization, while in Appendix E we will present the coherent states one. 

Analogously to what we did in (\ref{35-1}) let us perform a partial Fourier transform in order to go from the standard $\langle q,p,c^q,c^p|$ basis of the CPI to the $\langle \lambda_q,p,\bar{c}_q,c^p|$ one. In this new basis the transition amplitude is related to the old one in the following manner:
\begin{eqnarray}
&& \displaystyle \langle \lambda_q,p,\bar{c}_q,c^p,t| \lambda_{q_0},p_{\ssc 0},\bar{c}_{q_0},c^{p_0},t_{\ssc 0} \rangle= \label{46-1} \\
&& \displaystyle = \int dqdq_{\ssc 0} dc^q dc^{q_0} e^{-i\lambda_qq+\bar{c}_qc^q}
\langle q,p,c^q,c^p,t| q_{\ssc 0},p_{\ssc 0},c^{q_0},c^{p_0},t_{\ssc 0}\rangle 
e^{i\lambda_{q_0}q_{\ssc 0}-\bar{c}_{q_0}c^{q_0}}. \nonumber 
\end{eqnarray}
Replacing, on the RHS of the formula above, the kernel $\langle q,p,c^q,c^p,t|q_{\ssc 0},p_{\ssc 0},c^{q_0},
c^{p_0},t_{\ssc 0} \rangle$ with its path integral expression given in (\ref{34-1}), and using a properly defined discretized form of this path integral we get:
\begin{eqnarray}
&& \displaystyle \langle \lambda_q,p,\bar{c}_q,c^p,t|\lambda_{q_0},p_{\ssc 0},\bar{c}_{q_0},c^{p_0},t_{\ssc 0}\rangle = \label{47-1} \\
&& \displaystyle = \int {\mathscr D}Q {\mathscr D}^{\prime\prime}P \, \exp i\int_{t_0}^t id\tau d\theta d\bar{\theta} \left\{ L[\Phi]-\frac{d(QP)}{d\tau} \right\}. \nonumber
\end{eqnarray}
Differently from the case of the $\langle q,\lambda_p,c^q,\bar{c}_p|$ states, the surface terms coming from the partial Fourier transform in (\ref{46-1}) and those coming from (\ref{34-2}) do not combine to cancel against each other.
Nevertheless, as indicated explicitly, the surface terms that remain in (\ref{47-1}), i.e. $\displaystyle \frac{d(QP)}{dt}$, can be written in terms of the superphase space variables $Q$ and $P$, as we will prove in detail in Appendix D. Also the states appearing in (\ref{47-1}), i.e. $| \lambda_q, p, \bar{c}_q,c^p\rangle$, can be formally written in terms of the supervariables. In fact we should note that the variables $(\lambda_q,p,\bar{c}_q,c^p)$ entering these states are exactly the components of the $P$ of (\ref{36-2}). 
Analogously to what we did in (\ref{37-1}) for the state $|Q\rangle$, we could define the state $|P\rangle$ as the common eigenstate of the operators $\hat{p}$, $\hat{\lambda}_q$, $\hat{\bar{c}}_q$ and $\hat{c}^p$:
\begin{displaymath}
\begin{array}{l}
\hat{p}|P\rangle =p|P\rangle, \medskip \\
\hat{\bar{c}}_q|P\rangle =\bar{c}_q|P\rangle,
\end{array} \qquad \begin{array}{l}
\hat{\lambda}_q|P\rangle =\lambda_q|P\rangle, \medskip\\
\hat{c}^p |P\rangle =c^p|P\rangle. \label{48-1}
\end{array}
\end{displaymath}
The equations above can also be written in a compact form as $\hat{P}(\theta,\bar{\theta})|P\rangle =P(\theta,\bar{\theta})|P\rangle$. It is then natural to identify the state $|P\rangle$ with the state $|\lambda_q, p, \bar{c}_q,c^p\rangle$ appearing in (\ref{47-1}),\footnote{Using the completeness relations which arise from these $|P\rangle$ states and the $|Q\rangle$ ones of Eq. (\ref{37-2}) it is possible to build, via the standard Trotter formula, the path integral of the CPI directly in the superfield (or superphase variables) form. It is also possible to derive the graded commutators (\ref{14-2}) via generalized commutation relations among the superfields. Further details can be found in Appendix F.} which can be rewritten in the very compact form:
\begin{equation}
\displaystyle \langle P,t|P_{\ssc 0},t_{\ssc 0}\rangle = \int {\mathscr D}Q{\mathscr D}^{\prime \prime}P \, \exp i\int_{t_0}^t i d\tau d\theta d\bar{\theta} \left\{L[\Phi]-\frac{d(QP)}{d\tau}\right\}. \label{49-0} 
\end{equation}
Let us now remember the expression of the {\it quantum} transition amplitude in the momentum representation 
\begin{equation}
\displaystyle \langle p,t|p_{\ssc 0},t_{\ssc 0} \rangle = \int dq dq_{\ssc 0} \, e^{-ipq/\hbar} \langle q,t|q_{\ssc 0},t_{\ssc 0}\rangle e^{+ip_{\ssc 0}q_{\ssc 0}/\hbar}.
\label{49-1}
\end{equation}
By using a properly defined discretized form we get for (\ref{49-1}) the following expression\footnote{To make the reader understand how one can pass from the ${\mathscr D}^{\prime\prime} q {\mathscr D}p$ measure to the ${\mathscr D}q{\mathscr D}^{\prime\prime}p$ one in Appendix D we report the discretized derivation of this formula which can anyhow be found in the literature, see for example \cite{kleinert}.}
\begin{equation}
\displaystyle  \langle p,t| p_{\ssc 0}, t_{\ssc 0} \rangle = \int {\mathscr D}q{\mathscr D}^{\prime\prime} p \, \exp \frac{i}{\hbar} \int_{t_0}^t d\tau \, \left(L(q,p)-\frac{d(qp)}{d\tau} \right).
\label{49-3}
\end{equation}
At this point, looking at the {\it quantum} expression (\ref{49-3}) and at the {\it classical} one (\ref{49-0}), it is clear that the dequantization procedure that leads us from (\ref{49-3}) to (\ref{49-0}) is made of the same two rules which worked in the coordinate representation and which we wrote in italics in the lines below Eq. (\ref{38-2}). This confirms that our procedure works also in the momentum representation and that also in this case the same set of rules produces both the right evolution operator and the right representation. 
 
Before concluding this section, we would like to draw again the reader's attention to three crucial things, which made the whole procedure work as nicely as it did. The {\it first} one is that the classical weight $\int dt \widetilde{\cal L}$ and the quantum one $\int dt L$ belong, modulo surface terms, to the same multiplet. In fact if we expand $S[\Phi]= \int dt L[\Phi]$ in $\theta$ and $\bar{\theta}$ we get:
\begin{equation}
\displaystyle S[\Phi] = \int dt L(\varphi) + \theta {\cal T}(\varphi,\lambda,c,\bar{c})
+\bar{\theta} {\cal V} (\varphi, \lambda,c,\bar{c}) +i\theta \bar{\theta} \left( \int dt \widetilde{\cal L} (\varphi,\lambda,c,\bar{c}) +\textrm{s.t.} \right), \label{51-1} 
\end{equation}
 where the functions ${\cal T}$ and ${\cal V}$ are the analog of the $N$ and $\bar{N}$ which appeared in (\ref{29-2}). It is not important to write down the explicit form of ${\cal T}$ and ${\cal V}$ but to note that in (\ref{51-1}) the weights entering respectively the QPI, i.e. $\int dt L(\varphi)$, and the CPI, i.e. $\int dt \widetilde{\cal L}(\varphi, \lambda, c,\bar{c})$, belong to the same multiplet (modulo surface terms). Somehow we can say that ``{\it God has put CM and QM in the same multiplet (modulo surface terms)}". 
 
 The {\it second thing} to note is that in the statement in italics and between quotation marks written above we can even drop the ``modulo surface terms" part of the sentence. In fact those surface terms are crucial because, combined with the extra surface terms coming from the partial Fourier transforms of Eq. (\ref{35-1}) or (\ref{46-1}), they give exactly the classical weight that goes into the quantum one by the single procedure of sending $\theta, \bar{\theta} \to 0$. 
 
 The {\it third crucial thing} to note is that the partial Fourier transform of Eq. (\ref{35-1}) or (\ref{46-1}), which produces the extra surface terms exactly needed to implement the procedure above, it produces also the right classical states that go into the quantum ones by the same process of sending $\theta, \bar{\theta} \to 0$. This set of ``incredible coincidences" works both for the momentum and the coordinate representations and, as shown in Appendix E, also for the coherent states ones. 
 
The last issue that we want to touch is the one of ordering ambiguities. We know that there are no ordering problems in CM, at least in its standard formulation, while they are present in QM. This issue is at the basis of the fact that the quantization is not unique, i.e. starting from the same classical Hamiltonian we can end up in many different  ones at the quantum level according to the ordering we choose. Is this ambiguity resolved by our method of quantization? The answer is no. In fact CM, even if formulated \`a la KvN (or CPI), presents no ordering ambiguities as proved in detail in Appendix G. A detailed proof of this fact was needed because in principle the KvN (and CPI) formulation contains operators like $\hat{\varphi}^a$, $\hat{\lambda}_a$, which do not commute and so it could have presented ordering ambiguities. Having no ordering problem at the CPI level we conclude that these problems are injected into QM, not by the non-commuting operators of the CPI, but by the limiting procedure of sending $\theta,\bar{\theta} \to 0$. This limit in fact changes the kinetic term in the CPI from $\Phi^p\dot{\Phi}^q$ to $p\dot{q}$ which in turn implies the non-commutativity of $p$ and $q$ at the quantum level. 
We can conclude by saying that our quantization procedure does not solve the lack of uniqueness present in the standard quantization process.

\section{Generating Functionals and Dyson-Schwinger Equations}

The correspondence that we established between the transition amplitudes (\ref{38-1}) and (\ref{38-2}) can be extended also to more general objects, like the ``analog" of the generating functionals $Z[J]$ of quantum field theory.  The object we have in mind can be defined as follows
\begin{equation}
\displaystyle Z_{\ssc \textrm{QM}} \left( [J_a(t)];q,q_{\ssc 0},t-t_{\ssc 0} \right) = \int^q_{q_0} {\mathscr D}^{\prime\prime}q{\mathscr D}p \; \exp \left[\frac{i}{\hbar}\int_{t_0}^t d\tau \, \left( L[\varphi]+J_a\varphi^a \right)\right]. \label{53-1} 
\end{equation} 
It is not the vacuum-vacuum functional of quantum field theory but a {\it functional} of some external currents $J_a$ and a {\it function} of the initial $(q_{\ssc 0},t_{\ssc 0})$ and final $(q,t)$ points which generalizes the transition amplitude (\ref{38-1}). 

The classical analog of $Z_{\ssc \textrm{QM}}$ is the generalization at the level of generating functional \cite{due} of the expression (\ref{38-2}):
\begin{eqnarray}
\displaystyle && Z_{\ssc \textrm{CM}} \left( [\mathbb{J}_a(t,\theta,\bar{\theta})]; Q,Q_{\ssc 0},t-t_{\ssc 0} \right)=  \label{54-1} \\
\displaystyle && = \int_{Q_0}^Q {\mathscr D}^{\prime\prime} Q {\mathscr D}P \exp \left[ i\int_{t_0}^t  id\tau d\theta d\bar{\theta} \;L[\Phi] +i \int_{t_0}^t d\tau  \, (J_{\varphi^a}\varphi^a+J_{\lambda_a}\lambda_a +J_{c^a}c^a +\bar{c}_aJ_{\bar{c}_a}) \right]. \nonumber
\end{eqnarray}
The $8n$ currents ($J_{\varphi^a},J_{\lambda_a},J_{c^a},J_{\bar{c}_a}$) can be grouped together in a supercurrent ${\mathbb J}_a$ defined as 
\begin{displaymath}
\displaystyle {\mathbb{J}}_a\equiv\omega_{ab} J_{\lambda_b}-i\theta \omega_{ab}J_{\bar{c}_b}
+i\bar{\theta}J_{c^a}-i\bar{\theta}\theta J_{\varphi^a}. \label{54-2}
\end{displaymath}
Using this expression, the terms of (\ref{54-1}) depending on the currents can be written in a more compact form as
\begin{displaymath}
\displaystyle \int_{t_0}^t d\tau \left( J_{\varphi^a}\varphi^a+J_{\lambda_a}\lambda_a+J_{c^a}c^a+\bar{c}_aJ_{\bar{c}_a} \right)=
i\int_{t_0}^t d\tau d\theta d\bar{\theta} ({\mathbb{J}}_a \Phi^a).
\end{displaymath}
In this manner the $Z_{\ssc \textrm{CM}}$ of (\ref{54-1}) assumes the following compact expression 
\begin{equation}
\displaystyle Z_{\ssc \textrm{CM}} \left( [{\mathbb{J}}_a(t,\theta, \bar{\theta}) ]; Q,Q_{\ssc 0},t-t_{\ssc 0} \right)= \int_{Q_0}^Q {\mathscr D}^{\prime\prime}Q {\mathscr D}P 
\exp \left[ i\int_{t_0}^t id\tau d\theta d\bar{\theta} \,\left\{ L[\Phi]+{\mathbb{J}}_a\Phi^a \right\} \right],
\label{55-1}
\end{equation}
which can be compared to the quantum one of (\ref{53-1}) 
\begin{equation}
\displaystyle Z_{\ssc \textrm{QM}} ([J_a(t)]; q,q_{\ssc 0},t-t_{\ssc 0}) = \int_{q_0}^q
{\mathscr D}^{\prime\prime}q {\mathscr D}p \, \exp \left[ \frac{i}{\hbar} \int_{t_0}^t d\tau (L[\varphi]+J_a\varphi^a)\right]. \label{55-2}
\end{equation}
One immediately notices that the dequantization, i.e. passing from $Z_{\ssc \textrm{QM}}$ to $Z_{\ssc \textrm{CM}}$, is achieved by applying to the RHS of (\ref{55-2}) the following three rules
\begin{itemize}
\item[{\bf 1)}] $\displaystyle \int d\tau \; \longrightarrow \; i \hbar \int d\tau d\theta d\bar{\theta}$, 
\item[{\bf 2)}] $\displaystyle \varphi^a \; \longrightarrow \; \Phi^a$,
\item[{\bf 3)}] $\displaystyle J_a(t) \; \longrightarrow \; {\mathbb{J}}_a(t,\theta,\bar{\theta})$.
\end{itemize}  
There is the rule no. {\bf 3)}, besides the two old ones, because $Z$ is a functional of the current. It is clear that these rules work because $Z_{\ssc \textrm{CM}}$ and $Z_{\ssc \textrm{QM}}$ have the same functional form. This identity in the functional form has a further consequence. We know in fact that the generating functional $Z[J]$ satisfies a functional equation, known as Dyson-Schwinger equation \cite{Dyson}. Since both $Z_{\ssc \textrm{CM}}$ and $Z_{\ssc \textrm{QM}}$ have the same functional form we expect that they satisfy two Dyson-Schwinger equations with the same functional form. This is what we are going to prove.

We know \cite{Dyson} that 
the integral of a total derivative is zero and this works also in the functional case, so once these two operations are applied in sequence to a functional $F[\varphi]$ we get zero, i.e. 
\begin{displaymath}
\displaystyle \int {\mathscr D}\varphi \frac{\delta}{\delta \varphi} F[\varphi]=0. \label{57-00}
\end{displaymath}
If we choose $F[\varphi]$ to be $\displaystyle \exp \left[\frac{i}{\hbar}\left(S[\varphi]+\int dt J_a\varphi^a\right)\right]$, where $\displaystyle S[\varphi]=\int dt L(\varphi)$, we get:
\begin{displaymath}
\displaystyle \int {\mathscr D} \varphi \frac{\delta}{\delta \varphi} \, \left[\exp \frac{i}{\hbar}\left(S[\varphi]+\int dt J_a\varphi^a\right)\right]=0,
\end{displaymath}
which is equivalent to:
\begin{equation}
\displaystyle \int {\mathscr D}\varphi \, \left[\frac{\delta S}{\delta \varphi}+J\right]
\exp\left[ \frac{i}{\hbar} \left(S[\varphi] +\int dt J_a\varphi^a \right)\right]=0. \label{57-0}
\end{equation}
Remembering that
\begin{equation}
\displaystyle -i\hbar \frac{\delta}{\delta J_a} \exp \left[ \frac{i}{\hbar} \int dt J_a \varphi^a \right]= \varphi^a \; \exp \left[ \frac{i}{\hbar} \int dt J_a \varphi^a \right]  \label{replacement}
\end{equation}
we can replace the fields $\varphi^a$ with $\displaystyle -i\hbar \frac{\delta}{\delta J_a}$ and rewrite (\ref{57-0}) as:
\begin{equation}
\displaystyle \left[\frac{\delta S}{\delta \varphi}\left(\textstyle{-i\hbar\frac{\delta}{\delta J}}\right)
+J(t) \right]\, Z_{\ssc \textrm{QM}}[J]=0, \label{57-1}
\end{equation}
where we have used the explicit expression of the generating functional (\ref{53-1}).
Eq. (\ref{57-1}) is the Dyson-Schwinger equation satisfied by the generating functional $Z_{\ssc \textrm{QM}}[J]$ and it can also be rewritten as:
\begin{equation}
\displaystyle \left[ \partial_t \left( -i\hbar \frac{\delta}{\delta J_a(t)} \right) -\omega^{ab}\partial_bH \left(\textstyle{-i\hbar \frac{\delta}{\delta J(t)}}\right) +\omega^{ab}J_b\right] Z_{\ssc \textrm{QM}}[J_a]=0, \label{58-1}
\end{equation}   
where the combination of functional derivatives present on the LHS of (\ref{58-1}) 
comes from the term $\displaystyle \frac{\delta S}{\delta \varphi}$ in (\ref{57-1}), which in our case is $\partial_t\varphi^a-\omega^{ab}\partial_bH(\varphi)$, and by replacing $\varphi$ with $\displaystyle -i\hbar \frac{\delta}{\delta J}$, as dictated by the rule (\ref{replacement}) of the Dyson-Schwinger equation. 

Performing the same steps for the {\it classical} generating functional $Z_{\ssc \textrm{CM}}[{\mathbb{J}}]$ written in (\ref{55-1}) we get
\begin{equation}
\displaystyle \left[\partial_t\left( -\frac{\delta}{\delta {\mathbb{J}}_a(t,\theta,\bar{\theta})}
\right) -\omega^{ab}\partial_bH \left(\textstyle{-\frac{\delta}{\delta {\mathbb{J}}(t,\theta,\bar{\theta})}}\right)+\omega^{ab}{\mathbb{J}}_b(t,\theta, \bar{\theta})\right]Z_{\ssc \textrm{CM}}[\mathbb{J}]=0. \label{58-2} 
\end{equation}
The combination of superfunctional derivatives present on the LHS of (\ref{58-2}) is obtained from the LHS of the super-equations of motion $\displaystyle \partial_t\Phi^a-\omega^{ab}\frac{\partial H}{\partial \Phi^b}(\Phi)=0$
which, as proved in Ref. \cite{due}, are formally equivalent to the set of $8n$ equations of motion (\ref{20-1}). To get (\ref{58-2}) we must also remember that the definition of the functional derivative with respect to the supercurrent,
$\displaystyle \frac{\delta}{\delta \mathbb{J}_a(t,\theta,\bar{\theta})} \mathbb{J}_d(t,\theta^{\prime},\bar{\theta}^{\prime})=\delta_d^a\delta(\bar{\theta}^{\prime}-\bar{\theta})\delta({\theta}^{\prime}-\theta)$, implies that the analog of Eq. (\ref{replacement}) is given by:
\begin{eqnarray}
&& \displaystyle \frac{\delta}{\delta \mathbb{J}_a(t,\theta,\bar{\theta})} \left[\exp\left(i\int i dt^{\prime}d\theta^{\prime}d\bar{\theta}^{\prime} \, \mathbb{J}_d(t^{\prime},\theta^{\prime},\bar{\theta}^{\prime})\Phi^d(t^{\prime},\theta^{\prime},\bar{\theta}^{\prime})\right)\right]=\nonumber \medskip \\
&& =-\Phi^a(t,\theta,\bar{\theta})\left[\exp\left(i\int i dt^{\prime}d\theta^{\prime}d\bar{\theta}^{\prime} \, \mathbb{J}_d(t^{\prime},\theta^{\prime},\bar{\theta}^{\prime})\Phi^d(t^{\prime},\theta^{\prime},\bar{\theta}^{\prime})\right)\right]. \nonumber
\end{eqnarray}
The previous equation tells us that we can replace the superfield variable $\Phi(t,\theta,\bar{\theta})$ with the functional derivative 
$\displaystyle -\frac{\delta}{\delta \mathbb{J}(t,\theta,\bar{\theta})}$, as we did to get Eq. (\ref{58-2}).

Eqs. (\ref{58-2}) and (\ref{58-1}) prove that the {\it classical} and the {\it quantum} Dyson-Schwinger equations are formally very similar. One can pass from one to the other via the replacements 
\begin{itemize}
\item[{\bf 1)}] $\displaystyle J_a(t) \; \longrightarrow \; {\mathbb{J}}_a(t,\theta,\bar{\theta})$,
\item[{\bf 2)}] $\displaystyle -i\hbar \frac{\delta}{\delta J_a(t)} \; \longrightarrow \; -\frac{\delta}{\delta {\mathbb{J}}_a(t,\theta,\bar{\theta})}$.
\end{itemize}
The reader may think that rules {\bf 1)} and {\bf 2)} are in contradiction with each other because in {\bf 1)} there is no $\hbar$ while in {\bf 2)} an $\hbar$ makes its appearance. Actually  there is no contradiction because we should remember that the functional derivative
acts on  integrated things, so for example $\displaystyle \frac{\delta}{\delta \mathbb{J}_a(t,\theta,\bar{\theta})}$ acts on objects like $\displaystyle \int dt d\theta d\bar{\theta} \, F[\mathbb{J}_a(t,\theta,\bar{\theta})].$ We know that one of the dequantization rules is 
\begin{displaymath}
\displaystyle \int dt \, \longrightarrow \, i\hbar \int dt d\theta d\bar{\theta}.
\end{displaymath}
So if we pass from the functional derivation in $J(t)$ to the extended one:
\begin{displaymath}
 \displaystyle \frac{ \displaystyle \delta \int dt^{\prime} F[J(t^{\prime})]}{ \displaystyle \delta J(t)} \, \longrightarrow \,  \frac{ \displaystyle \delta \int dt^{\prime}d\theta^{\prime}d\bar{\theta}^{\prime}
F[\mathbb{J}(t^{\prime}, \theta^{\prime},\bar{\theta}^{\prime})]}{\displaystyle \delta \mathbb{J}(t,\theta,\bar{\theta})}=\frac{ \displaystyle \delta \left[ i\hbar \int dt^{\prime}d\theta^{\prime} d\bar{\theta}^{\prime}
F[\mathbb{J}(t^{\prime},\theta^{\prime},\bar{\theta}^{\prime})] \right]}{ \displaystyle i\hbar \,\delta \mathbb{J}(t,\theta,\bar{\theta})}
\end{displaymath}
we see that the correspondence between the functional derivatives is
$\displaystyle \frac{\delta}{\delta J(t)} \, \longrightarrow \, \frac{1}{i\hbar} \frac{\delta}{\delta \mathbb{J}(t,\theta,\bar{\theta})}$
which is equivalent to rule {\bf 2)}:
\begin{displaymath}
\displaystyle -i\hbar \frac{\delta}{\delta J(t)} \, \longrightarrow \, -\frac{\delta}{\delta \mathbb{J}(t,\theta,\bar{\theta})}.  \label{60-3}
\end{displaymath}

\section{Warnings on the Dequantization Rules}

The three dequantization rules that we have proposed up to now, and listed under Eq. (\ref{55-2}), need some further specifications that we are going to discuss in what follows.
\begin{itemize}
\item[{\bf A)}] If we take the path integral expression for the quantum transition amplitude (\ref{38-1}), and perform explicitly the functional integration (when this can be done exactly) we will get a function $K_{\ssc \textrm{QM}}(q,q_{\ssc 0};t-t_{\ssc 0})$ of the initial and final configurations $q_{\ssc 0}$ and $q$, and of the interval of time $(t-t_{\ssc 0})$:
\begin{displaymath}
\displaystyle \langle q,t|q_{\ssc 0},t_{\ssc 0}\rangle =K_{\ssc \textrm{QM}}(q,q_{\ssc 0};t-t_{\ssc 0}). \label{61-1}
\end{displaymath}
Analogously if we do the same for the classical transition amplitude (\ref{38-2}) we will obtain something like: 
\begin{displaymath}
\displaystyle \langle Q,t|Q_{\ssc 0},t_{\ssc 0}\rangle =\widetilde{K}_{\ssc \textrm{CM}}(q,\lambda_p,c^q,\bar{c}_p,q_{\ssc 0},\lambda_{p_0},c^{q_0},\bar{c}_{p_0};t-t_{\ssc 0}),
\label{62-2}
\end{displaymath}
where $\widetilde{K}_{\ssc \textrm{CM}}$ is a function of the components of the initial and final superfields $Q_{\ssc 0}$ and $Q$.
If we now naively apply one of the dequantization rules which says: ``Replace $q$ with $Q$ in order to pass from QM to CM", we expect that
\begin{equation}
\displaystyle K_{\ssc \textrm{QM}} (Q,Q_{\ssc 0}; t-t_{\ssc 0})=\widetilde{K}_{\ssc \textrm{CM}}.
\label{62-3}
\end{equation}
This is not so for a simple reason: the classical transition amplitude $\widetilde{K}_{\ssc \textrm{CM}}$ on the RHS of Eq. (\ref{62-3}) is equivalent to the kernel of propagation (\ref{35-1}), which does not depend on the Grassmann partners of time $\theta$ and $\bar{\theta}$, while the $K_{\ssc \textrm{QM}}(Q,Q_{\ssc 0})$ on the LHS of (\ref{62-3}) would depend on $\theta$, $\bar{\theta}$ via the supervariables $Q$ and $Q_{\ssc 0}$.

This is a first indication that we should not apply the dequantization rule mentioned above in an indiscriminate way. The replacement $q \to Q$ or $\varphi \to \Phi$ works only if we apply it to the functional integration measure and to the weight entering the path integral, or in general to any {\it functional} expression, like it happens for the Dyson-Schwinger equation. The replacement does not work if we do it in a {\it function}. 

\item[{\bf B)}] A further example of this fact comes from the comparison between the quantum commutators
\begin{equation}
\displaystyle \left[ \hat{\varphi}^a(t),\hat{\varphi}^b(t) \right]=i\hbar \omega^{ab}
\label{63-1}
\end{equation}
and the classical ones (\ref{14-2}) which, as shown in Appendix F, can be easily written in terms of superphase space variables as 
\begin{equation}
\displaystyle \left[ \hat{\Phi}^a(t,\theta,\bar{\theta}),\hat{\Phi}^b(t, \theta^{\prime},\bar{\theta}^{\prime}) \right]=\omega^{ab}\delta(\bar{\theta}-\bar{\theta}^{\prime})
\delta(\theta-\theta^{\prime}). \label{63-2}
\end{equation}
We immediately see that, replacing in (\ref{63-1}) $\hat{\varphi}^a$ with $\hat{\Phi}^a$, we do not end up in (\ref{63-2}). So the simple dequantization rule, which says ``replace $\varphi$ with $\Phi$", does not work also in this case. Again we have to apply it, but at a {\it functional} level. So we should start from the path integral (or functional) expression that is responsible for producing, at the operatorial level, the relation (\ref{63-1}). That expression is the kinetic part of the quantum path integral (\ref{38-1}) 
\begin{displaymath}
\displaystyle \int {\mathscr D}^{\prime\prime}q{\mathscr D}p \, \exp \frac{i}{\hbar}
\int_{t_0}^t d\tau \Bigl[ p\dot{q}-(\cdots) \Bigr].
\label{64-1}
\end{displaymath}
Applying the dequantization rules: $\varphi^a \to \Phi^a$ and $\int d\tau \to i \hbar \int d\tau d\theta d\bar{\theta}$ at this {\it functional} level, we get
\begin{displaymath}
\displaystyle \int {\mathscr D}^{\prime\prime} Q {\mathscr D}P \, \exp i\int_{t_0}^t i d\tau d\theta d\bar{\theta} \left[ P\dot{Q}- (\cdots) \right] \label{64-2}
\end{displaymath}
and from this expression we can derive, as proved in Appendix F, the commutators (\ref{63-2}). So we can pass from (\ref{63-1}) to (\ref{63-2}), but not by naively doing the replacement $\varphi^a \to \Phi^a$ at the level of the {\it functions} of (\ref{63-1}). We have to do that replacement, as in the case {\bf A)} analyzed before, at the level of the path integral (or {\it functional}) which produces the expression (\ref{63-1}).

\item[{\bf C)}] A third example, which tells us that we have to apply the dequantization rules only at the {\it functional} level, comes from looking at the observables of the quantum theory.
We know that in QM the observables are given by the Hermitian operators of $\hat{p}$ and $\hat{q}$. Let us indicate them as $O(\hat{\varphi}^a)$:
\begin{displaymath}
\displaystyle \left[O(\hat{\varphi}^a)\right]^{\dagger}=O(\hat{\varphi}^a). \label{65-1}
\end{displaymath}
If we naively apply the dequantization rule $\hat{\varphi}^a \rightarrow \hat{\Phi}^a$ we would get that the classical observables are given by 
$O(\hat{\Phi}^a)$.
These are objects that would depend on $\theta$ and $\bar{\theta}$, while we know that the classical observables are just functions of $p$ and $q$ and the generators of the canonical transformations at the CPI level are the Lie derivatives along the Hamiltonian vector fields associated with $O(\varphi^a)$. How can we recover these objects by going through our dequantization rules? Again the trick is to apply the rules at the {\it functional} level.

How can we go from a quantum observable to some ``{\it functional}" expression? Well, we can associate to every Hermitian operator ${O}(\hat{\varphi})$ a unitary one, defined as 
\begin{displaymath}
\displaystyle \hat{U}_{{O}}(\alpha)\equiv \exp \left[-\frac{i}{\hbar}O(\hat{\varphi}) \alpha\right],
\end{displaymath}
where $\alpha$ is a real parameter labeling the transformation generated by $O(\hat{\varphi})$. We can then sandwich this operator among two states $\langle q|$ and $| q_{\ssc 0}\rangle$:
\begin{equation}
\langle q| \exp \left[-\frac{i}{\hbar} {O}(\hat{\varphi})\alpha\right] |q_{\ssc 0}\rangle. \label{66-1}
\end{equation}
This expression is analog to the kernel of evolution where ${O}(\hat{\varphi})$ is replaced by ${H}(\hat{\varphi})$ and $\alpha$ is the interval of time $t$
\begin{equation}
\displaystyle \langle q| \exp \left[-\frac{i}{\hbar} H(\hat{\varphi}) t\right]|q_{\ssc 0}\rangle. \label{66-2}
\end{equation}
We know that Eq. (\ref{66-2}) has a path integral expression obtained by slicing the interval $t$ in $N$ parts and inserting the completeness relations. The same can be done for (\ref{66-1}) as follows:
\begin{eqnarray}
\displaystyle && \langle q| \exp \left[-\frac{i}{\hbar}{O}(\hat{\varphi})\alpha\right] |q_{\ssc 0}\rangle= \nonumber\\
\displaystyle && = \lim_{N\to \infty} \langle q| e^{-\frac{i}{\hbar} \, \hat{O}\frac{\alpha}{N}}
\int \frac{dp_{\ssc N}}{2\pi\hbar}|p_{\ssc N}\rangle \langle p_{\ssc N}|  \int dq_{\ssc N-1}|q_{\ssc N-1}\rangle \langle q_{\ssc N-1}|   e^{-\frac{i}{\hbar} \, \hat{O} \frac{\alpha}{N}} 
 \nonumber \\
\displaystyle && \quad \int \frac{dp_{\ssc N-1}}{2\pi\hbar} |p_{\ssc N-1}\rangle \langle p_{\ssc N-1}| \cdots \int dq_{\ssc 1}|q_{\ssc 1}\rangle \langle q_{\ssc 1}| e^{-\frac{i}{\hbar} \hat{O}\frac{\alpha}{N}} \int \frac{dp_{\ssc 1}}{2\pi\hbar}|p_{\ssc 1}\rangle \langle p_{\ssc 1}|q_{\ssc 0}\rangle. \nonumber
\end{eqnarray}
This leads to the following expression
\begin{displaymath}
\displaystyle \langle q|\exp \left[ -\frac{i}{\hbar} \, \hat{O}\alpha \right] |q_{\ssc 0}\rangle =\prod_{j=1}^{N-1} \int dq_j
\prod_{j=1}^{N} \int \frac{dp_j}{2\pi\hbar} \exp \left[\frac{i}{\hbar} {\mathcal O} \right], 
\end{displaymath}
where $\displaystyle {\mathcal O}= \sum_{j=1}^{N} \left[ p_j(q_j-q_{j-1})-\frac{\alpha}{N} O(q_j,p_j) \right]$, which formally can be written in a path integral form as
\begin{equation}
\displaystyle \langle q| \exp \left[-\frac{i}{\hbar} \, \hat{O}\alpha\right] |q_{\ssc 0}\rangle =\int_{q_0}^q {\mathscr D}^{\prime\prime} q{\mathscr D}p \exp \frac{i}{\hbar} \int_{0}^{\alpha}
d\bar{\alpha} \left[ p\frac{dq}{d\bar{\alpha}} -O(q,p) \right]. \label{67-2}
\end{equation}
In this way we have built a ``{\it functional}" expression from the operator $\hat{O}$. Now we can apply to the functional on the RHS of (\ref{67-2}) the following dequantization rules:
\begin{displaymath}
\left\{ \begin{array}{l}
\displaystyle \int d\bar{\alpha} \; \longrightarrow \; i\hbar \int d\bar{\alpha} d\theta d\bar{\theta}, \medskip\\
\varphi^a \; \longrightarrow \; \Phi^a.
\end{array}
\right.
\end{displaymath}
What we get is: 
\begin{equation}
\displaystyle \int_{Q_{\ssc 0}}^Q {\mathscr D}^{\prime\prime}Q {\mathscr D}P
\exp i\int_{0}^{\alpha} id\bar{\alpha} d\theta d\bar{\theta} \left[P\frac{dQ}{d\bar{\alpha}}
-O(\Phi)\right]. \label{68-1}
\end{equation}
Comparing this with (\ref{38-2}) it is clear that (\ref{68-1}) gives the classical transition amplitude $\langle Q,\alpha|Q_{\ssc 0},0\rangle$ which in turn can be written, analogously to (\ref{19-2}), as:
\begin{displaymath}
\displaystyle \langle Q, \alpha|Q_{\ssc 0},0\rangle =\langle Q| \exp -i\hat{\widetilde{O}}\alpha |Q_{\ssc 0}\rangle,
\end{displaymath}
where 
\begin{displaymath}
\displaystyle \hat{\widetilde{O}}=-i\omega^{ab}\partial_bO\partial_a-i\omega^{ab}\partial_b\partial_dO\hat{c}^d\frac{\partial}{\partial c^a}.
\end{displaymath}
This is the classical Lie derivative along the Hamiltonian vector field associated to $O(\varphi^a)$ \cite{otto} and is the analog of the $\hat{\widetilde{\cal H}}$ of (\ref{16-3}).
So this is the manner to build the {\it classical} analog of the transformations generated by the quantum operators ${O}(\hat{\varphi})$. Again we have applied our usual dequantization rules, but only at the ${\it functional}$ level.

\item[{\bf D)}] Another object on which we should be careful in naively applying our dequantization rules is the wave function. We had seen in Sec. 4 that the substitution $ q \to Q$ works at the level of abstract kets $|q\rangle$, in the sense that the states $|Q\rangle$ obtained via our substitution are really the states appearing in the classical transition amplitude (\ref{38-2}). Now if we try to apply this rule also to the wave functions
\begin{equation}
\displaystyle \langle q|\psi \rangle =\psi(q) \label{69-1}
\end{equation}
by substituting on the RHS above $q \to Q$, then we get
$\displaystyle \psi(q) \; \longrightarrow \; \psi(Q).$
Of course this $\psi(Q)$ does not belong to the ``classical" wave functions of KvN that we have introduced in (\ref{19-1}). $\psi(Q)$ in fact depends on $\theta$ and $\bar{\theta}$, while the KvN wave functions are just functions of $\varphi^a$ and $c^a$ or of $q$, $\lambda_p$ $c^q$ and $\bar{c}_p$, according to the representation we choose.
Actually also the ket $|Q\rangle$ did not depend on $\theta$ and $\bar{\theta}$, as explained in Sec. 4. In fact $|Q\rangle$ was defined as the eigenstate of the operator $\hat{Q}$
\begin{displaymath}
\displaystyle \hat{Q}(\theta,\bar{\theta})|Q\rangle =Q(\theta,\bar{\theta})|Q\rangle, \label{70-1}
\end{displaymath}
so the dependence on $\theta, \bar{\theta}$ was contained in the operator $\hat{Q}$ and in the eigenvalue $Q$ but not in $|Q\rangle$, which, as explained in Sec. 4, could be identified with $|q,\lambda_p,c^q,\bar{c}_p\rangle$. This means that the set of wave functions of the operatorial theory lying behind the CPI can be written as
\begin{displaymath}
\displaystyle \langle Q|\psi \rangle =\langle q,\lambda_p,c^q,\bar{c}_p|\psi\rangle=\psi(q,\lambda_p,c^q,\bar{c}_p). \label{70-2}
\end{displaymath}
This confirms that
the substitution $q \to Q$ cannot be applied to both sides of Eq. (\ref{69-1}) since the correct association is given by:
\begin{eqnarray}
\langle q|\psi \rangle &=& \psi(q) \nonumber \\
& \Downarrow & \nonumber\\
\langle Q|\psi \rangle &=& \psi(q,\lambda_p,c^q,\bar{c}_p) \neq \psi(Q). \nonumber
\end{eqnarray}

\item[{\bf E)}] The last warning we want to issue on our dequantization rules regards the extension in the base space $t\, \longrightarrow \, (t,\theta,\bar{\theta})$, given by the rule:
\begin{displaymath}
\displaystyle \int dt \; \longrightarrow \; i\hbar \int dt d\theta d\bar{\theta}. \label{73-1}
\end{displaymath}
We have seen that this rule works in all the cases we have treated and that even the functional derivative with respect to the current $J$ is strictly related to this rule. The reader may wonder that, if we extend the integration and the whole base space from $t$ to $(t,\theta,\bar{\theta})$ then we should extend also the 
derivative with respect to $t$, i.e. $d/dt$ to a sort of combined derivative, like for example the covariant one \cite{nove}-\cite{quindici}. This actually does not happen, as it is clear from the weight in (\ref{35-2}), which contains in $L[\Phi]$ the normal time derivative $\displaystyle \int dt d\theta d\bar{\theta} (P\dot{Q}+\cdots )$, or from the Dyson-Schwinger equation (\ref{58-2}) which also contains the time derivative. The fact that the integration gets extended, while the derivation is not, needs for sure further study. Besides this technical point, we feel that further work is needed to better understand the {\it physics} behind our dequantization procedure.
\end{itemize}

\section{Conclusions and Outlooks}

We will not summarize here what we did in this long paper and in its appendices. We will only write down the dequantization rules, which are now in their complete form, and next we will outline some physical ideas that we are pursuing at the moment. 

We can summarize our dequantization procedure as follows. In order to pass from QM to CM one must apply the following rules:
\begin{itemize}
\item[{\bf 1)}] Given a QM object, in case it is not already written in functional form, build from it a {\it functional} expression, which either gives the action of that object on the states, or from which the object itself can be derived;
\item[{\bf 2)}] Next, in that functional expression perform the following replacements:
\begin{itemize}
\item[{\bf A)}] the time integration with a proper supertime integration:
\begin{displaymath}
\displaystyle \int dt \; \longrightarrow \; i\hbar \int dt d\theta d\bar{\theta},
\end{displaymath}
\item[{\bf B)}] the phase space variables with the superphase space variables:
\begin{displaymath}
\displaystyle \varphi^a(t) \; \longrightarrow \; \Phi^a(t,\theta,\bar{\theta}),
\end{displaymath}
\item[{\bf C)}] and the functional derivatives with respect to the external currents $J(t)$ with those done with respect to the supercurrents $\mathbb{J}(t,\theta,\bar{\theta})$: 
\begin{displaymath}
\displaystyle i\hbar \frac{\delta}{\delta J(t)} \; \longrightarrow \; \frac{\delta}{\delta {\mathbb{J}}(t,\theta, \bar{\theta})}. 
\end{displaymath}
\end{itemize}
\end{itemize}
As we repeatedly said these rules are very {\it geometrical} (and beautiful in our opinion), but we feel that we should try to better understand the {\it physics} behind them. In what follows we will briefly and qualitatively present some of our ideas on this issue. We apologize in advance with the reader if some of these ideas may sound to him too vague or too farfetched and not yet technically very precise. We want only to convey the flavor of what may lie behind our geometrical construction.

The first issue of physics we want to understand is the physical meaning of the Grassmann partners of time $\theta$ and $\bar{\theta}$. We know that the Grassmann partners $c^a$ of the phase space variables $\varphi^a$ can be identified with the forms $d\varphi^a$ or the Jacobi fields $\delta \varphi^a$ between two classical trajectories \cite{due}. It seems then ``natural" to identify $\theta$ with $dt$ or $\Delta t$ that means with the infinitesimal of time. It is actually not so, as can be understood from dimensional reasons. From Appendix B and particularly formula (\ref{2-B}) it seems more appropriate to identify 
\begin{equation}
\Delta t \sim \theta \bar{\theta}. \label{77-1}
\end{equation}
Some other reasoning, rather approximate and intuitive, contained in Appendix H, seems to confirm formula (\ref{77-1}). 

Let us now remember our procedure of quantization, that means how to pass from (\ref{38-2}) to (\ref{38-1}). We said that it corresponds to taking the limit of $\theta,\bar{\theta} \to 0$. This can be technically achieved by inserting into the weight $\int dt d\theta d\bar{\theta} L[\Phi]$ the expression $\delta(\bar{\theta})\delta({\theta})/\hbar$, which is actually equal to $\bar{\theta} {\theta}/\hbar$, and also, using (\ref{77-1}), to $\Delta t/\hbar$. This tells us that we are inserting into CM the constraint 
\begin{equation}
\displaystyle \Delta t \sim \hbar. \label{77-2}
\end{equation}
This is somehow equivalent to the crucial feature of QM which says that \cite{sakita} 
\begin{equation}
\displaystyle (\Delta q)^2 \sim \Delta t \; \hbar \label{77-3}
\end{equation}
and which indicates the universal Wiener process present in QM. Now, as $\Delta q \sim \Delta t$ if $c=1$, we get from (\ref{77-3}) that $(\Delta t)^2 \sim \Delta t \, \hbar$, which is exactly (\ref{77-2}). So, summarizing, we can say that, by introducing the term $\delta(\bar{\theta})\delta({\theta})/\hbar$, we are injecting into the weight of CM the crucial feature of QM given by relation (\ref{77-3}). This gives a ``rough" physical explanation of the reasons why our geometrical dimensional reduction of supertime, achieved via $\theta, \bar{\theta} \to 0$, injects the right quantum mechanical ingredient. This analysis needs further refinement and a more precise technical study but we feel that, even in this rough form, it encapsulates the physics behind our process.

Some other ideas, which may convey the feeling of what is going on, are the following. We know \cite{jackiw} that time can be turned into an inverse temperature. Maybe something similar can be done with the Grassmann time. Then the limit $\theta,\bar{\theta} \to 0$, with which we get QM, may indicate that the ``temperature" associated to the Grassmann time is actually going to infinity in QM. This would mean that in QM there is a sort of very high (infinite) Grassmann temperature. This infinite temperature may be responsible for the zero point energy and explain why it cannot be turned off in QM. 

If the reader is not pleased with this concept of Grassmann temperature as inverse of the Grassmann time, we could propose the following other picture. A quantity that is the inverse of time is the frequency $\nu_t$. Maybe we can define also a frequency $\nu_{\theta}$ associated to the Grassmann time $\theta$. Then the $\theta, \bar{\theta} \to 0$ limit, with which we get QM, may indicate that the associated frequency $\nu_{\theta}$ is going to infinity. Of course $\nu_{\theta}$, like the temperature in the previous picture, is a Grassmann quantity and to say that we send it to infinity has to be properly defined. We are well aware that
these are all very qualitative and vague ideas but they might be the starting points to get some new view on QM.
Well, the first problem that CM faced was the black body radiation. There it appeared  that CM was giving the wrong results when $\nu_t \to \infty$ and we had to pass to QM which, in our picture, would mean $\nu_{\theta} \to \infty$. So somehow both frequencies must be taken to the same limit if we want the correct theory. Maybe this is a signal of a symmetry linking the two frequencies, like there is a symmetry linking $t$ and $\theta$ \cite{due}. It is as if God did not want this symmetry to be broken and forced us, when we send $\nu_t \to \infty$, to do the same with $\nu_{\theta}$.

The last idea we want to flash at the reader is the following. Up to now we have passed from QM to CM or from CM to QM by doing something {\it by hand}: either inserting a $\delta(\bar{\theta})\delta({\theta})/\hbar$ in the weight of (\ref{38-2}) to pass from CM to QM, or extending by hand the integration $\displaystyle  \int dt \longrightarrow i\hbar \int dt d\theta d\bar{\theta}$ and the phase space variables $\varphi^a \to \Phi^a$ to pass from QM to CM. We would like to find a manner to realize the same steps in a more natural way. An idea we have in mind is to study the various symmetries present in the CPI. Some of these symmetries were already discovered in Ref. \cite{due} and presented in Sec. 3. They are BRS-like and supersymmetry-like invariances plus some other ones, like ``ghost" number and ``ghost" conjugation. We feel that these are not the only universal symmetries present in the CPI. One idea is to make the $\widetilde{\cal L}_{\ssc \textrm{CPI}}$ invariant under a local supertime reparametrization of the form: 
\begin{equation}
\left\{ \begin{array}{l}
t^{\prime}=t^{\prime}(t,\theta,\bar{\theta}) \medskip\\
\theta^{\prime}=\theta^{\prime}(t,\theta,\bar{\theta}) \label{82-1} \medskip \\
\bar{\theta}^{\prime}=\bar{\theta}^{\prime}(t,\theta, \bar{\theta})
\end{array}
\right.
\end{equation}
or even a ``field-dependent" reparametrization of the form 
\begin{equation}
\left\{ \begin{array}{l}
t^{\prime}=t^{\prime}(t,\theta,\bar{\theta};\Phi^a) \medskip\\
\theta^{\prime}=\theta^{\prime}(t,\theta,\bar{\theta};\Phi^a) \label{82-2} \medskip \\
\bar{\theta}^{\prime}=\bar{\theta}^{\prime}(t,\theta, \bar{\theta};\Phi^a).
\end{array}
\right.
\end{equation}
We call it ``field-dependent" because the target space variables $\Phi^a$ are like fields with base space $(t,\theta,\bar{\theta})$. Let us call the Lagrangian invariant under (\ref{82-2}) as $\widetilde{\cal L}_{\textrm{inv}}$. The ``superdiffeomorphism"
invariance (\ref{82-1})-(\ref{82-2}) will require the introduction of a supermetric so that the ``super-interval" of time will be given by something like:
\begin{displaymath}
\displaystyle \Delta= g_{tt}(t_{\ssc 1}-t_{\ssc 2})^2 +g_{t\theta}(t_{\ssc 1}-t_{\ssc 2})(\theta_{\ssc 1}-\theta_{\ssc 2})+g_{t\bar{\theta}}(t_{\ssc 1}-t_{\ssc 2})(\bar{\theta}_{\ssc 1}-\bar{\theta}_{\ssc 2})+g_{\theta \bar{\theta}}(\theta_{\ssc 1}-\theta_{\ssc 2})(\bar{\theta}_{\ssc 1}-\bar{\theta}_{\ssc 2}).
\end{displaymath}
Of course the symmetries (\ref{82-1}) and (\ref{82-2}) must be gauge-fixed once we build  a path integral for $\widetilde{\cal L}_{\textrm{inv}}$. Maybe the old $\widetilde{\cal L}_{\ssc \textrm{CPI}}$ could be identified with the $\widetilde{\cal L}_{\textrm{inv}}$, once we choose a particular gauge-fixing. Also the weight of QM, $L(\varphi)$, could be a particular ``gauge-fixed" version of $\widetilde{\cal L}_{\textrm{inv}}$, in particular it could be the one for which 
\begin{displaymath}
g_{t\theta}=g_{t\bar{\theta}}=g_{\theta\bar{\theta}}=0. \label{84-1}
\end{displaymath}
The reason is that this choice of metric would take away every dependence on $\theta$, $\bar{\theta}$, like it must be in QM. So somehow we would get the limit of $\theta, \bar{\theta} \to 0$ by a proper choice of the metric or of the ``gauge-fixing". The picture that would emerge is that both CM and QM are different gauge-fixed versions of a unique theory given by the Lagrangian $\widetilde{\cal L}_{\textrm{inv}}$, see {\bf Fig. 1}.
\begin{figure}[h!]
\begin{center}

\rnode{A}{$\widetilde{\cal L}_{\textrm{inv}}$} 

\vspace{0.9cm} 
\hspace{9cm} {\bf Fig. 1} 
\vspace{0.1cm}

\rnode{B}{$\widetilde{\cal L}_{\ssc \textrm{CPI}}$} \hspace{3.7cm}
\rnode{C}{$L(\varphi)$}
\psset{nodesep=5pt, arrows=->}
\ncLine{A}{B}\ncLine{A}{C}
\psset{nodesep=5pt, arrows=<->}
\ncLine{B}{C}

\hspace{0.33cm} ? 
\vspace{1cm}

\rnode{D}{gauge-fixing (1)} \hspace{2cm}\rnode{E}{gauge-fixing (2)} 
\psset{nodesep=5pt, arrows=<-}
\ncLine{B}{D}\ncLine{C}{E}

\end{center}
\end{figure}

If QM and CM are different gauge-fixed
versions of the same theory then it should be possible to pass from one to the other via a gauge transformation. This would mean that QM and CM are equivalent theories and we know they are not. So why we cannot turn CM into QM by a gauge transformation? The reason may be that the gauge symmetry is affected by an anomaly and so it would not be possible to perform the gauge transformation which would ``rotate" $\widetilde{\cal L}_{\ssc \textrm{CPI}}$ into $L(\varphi)$. Usually anomalies appear in multidimensional field theories and not in point particle systems like ours. An exception is made by the ``large gauge transformations" \cite{weizmann} which could be anomalous also for systems with a finite number of degrees of freedom. So the anomaly we have in mind could be on the ``large gauge transformations" belonging to (\ref{82-1}). Another possibility is that the transformations of the type (\ref{82-2}) could present an anomaly even for systems with a finite number of degrees of freedom. These kind of ``field dependent" reparametrizations have not been studied much in the literature and we could find some surprises. 

Another reason why we feel an anomaly could affect the transformation (\ref{82-2}) is the following. There is a universal symmetry under which QM is ``anomalous" or at least non-invariant: it is the symmetry that rescales the overall action. In fact the presence in QM of $\hbar$, the quantum of action, would forbid a symmetry that rescales the action and we have indications that this symmetry belongs to the class of (\ref{82-2}). The reader may be rather puzzled by the picture we have outlined and ask how $\hbar$ would make its appearance from $\widetilde{\cal L}_{\ssc \textrm{CPI}}$. As everybody knows, anomalies are indicated by the non-invariance of the functional integration \cite{fujikawa}. The Jacobian that arises from the non-invariance of the measure must be calculated precisely and it usually needs a regularization procedure. We feel that $\hbar$ (or a cut off $\Delta$ on the action volume) is the regularizing parameter we need to calculate this determinant. Work is in progress on these ideas and we hope to be able to present it soon to the physics community. 

\section*{Acknowledgments}

This work has been supported by grants from the University of Trieste, MIUR and INFN of Italy and grants RFBR 03-02-16209
and RFBR NSh-1774.2003.2 of Russia. Two of us (E.G. and D.M.) warmly acknowledge many helpful discussions with M. Reuter. 

\newpage 
\begin{center}
{\LARGE\bf Appendices}
\end{center}

\appendix
\makeatletter
\@addtoreset{equation}{section}
\makeatother
\renewcommand{\theequation}{\thesection.\arabic{equation}}

\section{Appendix }

In this Appendix we will analyze the complex nature and the physical dimensions of the Grassmann variables $c$, $\bar{c}$, $\theta$, $\bar{\theta}$ entering the definition (\ref{21-1}) of the superfield $\Phi^a$:
\begin{displaymath}
\displaystyle \Phi^a=\varphi^a+\theta c^a +\bar{\theta} \omega^{ab}\bar{c}_b+i\bar{\theta}\theta \omega^{ab}\lambda_b.
\end{displaymath}

Let us consider the superfield $\Phi^a$ as an operator and impose the scalar product under which the Grassmann operators $\hat{c}$ and $\hat{\bar{c}}$ are Hermitian, see Ref. \cite{sette} for further details. Since the first component $\hat{\varphi}$ of the superfield $\hat{\Phi}$ is Hermitian we require the entire superfield operator to be Hermitian.
From 
\bea
&& \displaystyle \hspace{5cm} \hat{\Phi}^{a \dagger}=\hat{\Phi}^a  \Longrightarrow \medskip \nonumber\\ && \hat{\varphi}^a +\hat{c}^a\theta^* +\omega^{ab}\hat{\bar{c}}_b\bar{\theta}^*-i\omega^{ab}\hat{\lambda}_b\theta^*\bar{\theta}^* 
=\hat{\varphi}^a+\theta \hat{c}^a+\bar{\theta}\omega^{ab}\hat{\bar{c}}_b+i\bar{\theta}\theta \omega^{ab}\hat{\lambda}_b \nonumber
\eea
we easily get that the two Grassmann partners of time $\theta$ and $\bar{\theta}$ must be {\it imaginary}:
\be
\theta^*=-\theta, \qquad \quad \bar{\theta}^*=-\bar{\theta}. \label{imaginary}
\ee
Of course, other choices of the scalar product imply other conventions about the character of the variables $\theta$ and $\bar{\theta}$. For example, if we impose the symplectic scalar product \cite{sette}, under which the Hermiticity conditions among the Grassmann operators are given by:
\begin{displaymath}
\displaystyle (\hat{c}^a)^{\dagger} =i\omega^{ab}\hat{\bar{c}}_b,
\qquad (\hat{\bar{c}}_a)^{\dagger}=i\omega_{ab}\hat{c}^b,
\end{displaymath}
and we require the superfield to be Hermitian:
\bea
&& \displaystyle \hspace{5cm} \hat{\Phi}^{a \dagger}=\hat{\Phi}^a \, \Longrightarrow  \medskip \nonumber\\ && \hat{\varphi}^a+i\omega^{ab}\hat{\bar{c}}_b\theta^*+i\hat{c}^a\bar{\theta}^*-i\theta^*\bar{\theta}^*\omega^{ab}\hat{\lambda}_b =\hat{\varphi}^a+\theta \hat{c}^a +\bar{\theta} \omega^{ab}\hat{\bar{c}}_b+i\bar{\theta}\theta \omega^{ab}\hat{\lambda}_b
\nonumber 
\eea
we get the following relations among the Grassmann partners of time:
\begin{displaymath}
\displaystyle \bar{\theta}^*=i\theta, \qquad \theta^*=i\bar{\theta}.
\end{displaymath}
Finally, with the scalar product under which $\hat{c}^{a \dagger}=\hat{\bar{c}}_a$  \cite{sette}, it is impossible to have a Hermitian superfield operator. Anyhow, if we stick to the scalar product under which the operators $\hat{c}$ and $\hat{\bar{c}}$ are Hermitian, we can take the Grassmann partners of time $\theta$ and $\bar{\theta}$ to be imaginary, Eq. (\ref{imaginary}). With this choice it is easy to prove that the presence of the factor ``$i$" in Eq. (\ref{39-1}) is crucial to have a real integration measure:
\begin{displaymath}
\displaystyle (id\theta d\bar{\theta})^*=-id\bar{\theta}^*d\theta^*=-i(-d\bar{\theta})(-d\theta)= -i d\bar{\theta}d\theta =id\theta d\bar{\theta}.
\end{displaymath}
Not only, but with the scalar product under which the Grassmann operators $\hat{c}$ and $\hat{\bar{c}}$ are Hermitian the BRS and anti-BRS charge of Eq. (\ref{23-1}) are anti-Hermitian, i.e.
$\hat{Q}^{\dagger}=-\hat{Q}$ and $\hat{\bar{Q}}^{\dagger}=-\hat{\bar{Q}}$. Consequently, the operators $\exp \theta \hat{Q}$ and $\exp \hat{\bar{Q}} \bar{\theta}$, entering the connection between the Schr\"odinger and the Heisenberg pictures given by Eq. (\ref{connect}), are unitary. In fact the anti-Hermiticity of $\hat{Q}$, $\hat{\bar{Q}}$ and Eq. (\ref{imaginary}) imply:
\begin{displaymath}
\displaystyle [\exp (\theta \hat{Q})]^{\dagger}=\exp (\hat{Q}^{\dagger}\theta^*)=
\exp(\hat{Q}\theta)=\exp(-\theta \hat{Q}),
\end{displaymath}
which means that $\exp (\theta \hat{Q})$ is a unitary operator. A similar proof holds for $\exp(\hat{\bar{Q}}\bar{\theta})$. 
 
Another interesting point to discuss is the physical dimension of the Grassmann variables $\theta$ and $\bar{\theta}$. Let us start by considering that, in the definition (\ref{33-2}) of the superfields, $q$ and $\bar{\theta}\theta\lambda_p$ appear in the same multiplet. So, from the point of view of the physical dimensions, we will have:
\begin{displaymath}
[q]=[\bar{\theta}\theta  ][\lambda_p] \; \longrightarrow \; [\bar{\theta}\theta]=[q] [\lambda_p]^{-1}.
\end{displaymath}
Now $[p,\lambda_p]=i$, so the dimensions of $\lambda_p$ are just the inverse of the dimensions of $p$ and so we can derive that the product $\bar{\theta}\theta$ has the dimensions of an action:
\be
\displaystyle [\bar{\theta} \theta]=[q] [p] = [\hbar]. \label{action}
\ee
The dimensions of the product of the infinitesimals $d\theta d\bar{\theta}$ instead can be derived from the following equation:
\begin{displaymath}
\displaystyle \widetilde{\cal H}=i \int d\theta d\bar{\theta} \, H.
\end{displaymath}
As an operator $\widetilde{\cal H}$ is a derivative with respect to $t$, so it has the dimensions of the inverse of a time: $[\widetilde{\cal H}]=T^{-1}$ while $H$ is an energy, so its dimensions are: $[H]=ML^2T^{-2}$. This immediately implies that the dimensions of $d\theta d\bar{\theta}$ are given by:
\begin{displaymath}
\displaystyle [d\theta d\bar{\theta}]=[\widetilde{\cal H}]/[H]=M^{-1}L^{-2}T=L^{-1} (LMT^{-1})^{-1}=[q]^{-1}[p]^{-1}.
\end{displaymath}
From the previous equation we derive that the product $d\theta d\bar{\theta}$ has the dimensions of the inverse of an action:
\be
\displaystyle [d\theta d\bar{\theta}]=[\hbar]^{-1}.  \label{invaction}
\ee
The previous equation is consistent with the standard definition of the Grassmann integration, which through all this paper is the following one:
\begin{displaymath}
\displaystyle \int d\theta \,1=0, \qquad \int d\bar{\theta} \, 1=0, 
\qquad  \int d\theta \,\theta=1, \qquad \int d\bar{\theta} \, \bar{\theta}=1.
\end{displaymath}
Such definition implies that $[d\theta]=[\theta]^{-1}$ and $[d\bar{\theta}]=[\bar{\theta}]^{-1}$, i.e. the dimensions of the infinitesimal of a Grassmann variable are just the inverse of the dimensions of the Grassmann variable itself. This is perfectly consistent with Eq. (\ref{action}) and (\ref{invaction}).

Even if the dimensions of the product $\theta\bar{\theta}$ are uniquely determined, there is still an arbitrariness in the dimensions of the single variables $\theta$ and $\bar{\theta}$. In fact, from the expression of the superfields:
\begin{displaymath}
\displaystyle Q=q+\theta c^q+\bar{\theta}\bar{c}_p+i\bar{\theta}\theta \lambda_p, \medskip \label{prima} \qquad
\displaystyle P=p+\theta c^p-\bar{\theta}\bar{c}_q-i\bar{\theta}\theta \lambda_q, \end{displaymath}
and the fact that all the components of a superfield must have the same physical dimensions, we can derive that $q$ must have the same physical dimensions of $\theta c^q$ and $\bar{\theta}\bar{c}_p$, while $p$ must have the same physical dimensions of $\theta c^p$ and $\bar{\theta}\bar{c}_q$:
\begin{eqnarray}
&& \displaystyle [q]=[\theta c^q] =[\bar{\theta} \bar{c}_p] \label{prima1} \medskip \\
&& \displaystyle [p]=[\theta c^p] =[\bar{\theta} \bar{c}_q]. \label{seconda}
\end{eqnarray}
Nevertheless we cannot determine in a unique way the dimensions of the single fields $c^q$, $c^p$, $\bar{c}_q$ and $\bar{c}_p$. The only thing that we can derive, dividing (\ref{prima1}) by (\ref{seconda}), is that the following equations hold:
\be
\displaystyle \frac{[c^q]}{[c^p]}=\frac{[q]}{[p]}, \qquad \frac{[\bar{c}_q]}{[\bar{c}_p]}=\frac{[p]}{[q]}. \label{dimdue}
\ee
This means that only the ratio of the dimensions of the variables $c$ and the ratio of the dimensions of the variables $\bar{c}$ can be determined. 
It is quite easy to show that Eqs. (\ref{action}) and (\ref{dimdue}) guarantee the dimensional consistency of all the formulae that can be derived from the CPI. In any case, we must stress the fact that there is nothing in the theory that can fix the dimensions of the single Grassmann variables $c$ and $\bar{c}$. Consequently there is nothing in the theory that can fix the dimensions of the single Grassmann partners of time $\theta$ and $\bar{\theta}$. 

Anyhow since the Grassmann variables $c$ can be interpreted as the Jacobi fields or the first variations of the theory $c^a\approx \delta \varphi^a$, 
see \cite{quattro}-\cite{sette}, we could take the physical dimensions of $c^q$ to be identical to the physical dimensions of $q$ and the physical dimensions of $c^p$ to be identical to the physical dimensions of $p$. With these conventions we would have that $\theta$ is dimensionless while $\bar{\theta}$, in order to satisfy (\ref{action}), must have the dimensions of an action.

\newpage

\section{Appendix }

In this Appendix we will construct the various intervals of supertime which are invariant 
under susy \cite{undici}. By invariant under susy we mean an object which is left invariant by the transformations generated by the operator $\epsilon \hat{\cal Q}_{\ssc H}+\bar{\epsilon}\hat{\bar{{\cal Q}}}_{\ssc H}$
where $\epsilon$ and $\bar{\epsilon}$ are anticommuting parameters and $\hat{\cal Q}_{\ssc H}$ and $\hat{\bar{{\cal Q}}}_{\ssc H}$
are given in (\ref{25-1}). Such transformations are the following ones:
\be
\displaystyle \delta t =-\epsilon \bar{\theta}+\bar{\epsilon}\theta, \qquad \delta \theta =-\epsilon, \qquad \delta \bar{\theta}=\bar{\epsilon}. \label{biuno}
\ee
It is straightforward to check that the following interval, defined
between the two super-instants of time $(t_{\ssc 1},\theta_{\ssc 1},\bar{\theta}_{\ssc 1})$ and $(t_{\ssc 2},\theta_{\ssc 2},\bar{\theta}_{\ssc 2})$,
\be
\displaystyle S=t_{\ssc 2}-t_{\ssc 1}+\theta_{\ssc 2}\bar{\theta}_{\ssc 1}-\theta_{\ssc 1}\bar{\theta}_{\ssc 2} \label{2-B}
\ee
is left invariant by the transformations (\ref{biuno}). The careful reader should notice that in the definition of the super-interval of time $S$ we couple together $t$ with products of $\theta$ and $\bar{\theta}$. From what we said in the previous appendix, $\bar{\theta}\theta$ has the dimensions of an action and not of a time. Anyhow, there is no dimensional contradiction in Eq. (\ref{2-B}). In fact throughout this paper we have put the dimensional constant $\beta$, appearing in the original definition \cite{quattro} of the supersymmetry charges $Q_{\ssc H}$ and $\bar{Q}_{\ssc H}$, equal to one. If we included the term $\beta$ then we would get it in front of $\theta_{\ssc 2}\bar{\theta}_{\ssc 1}-\theta_{\ssc 1}\bar{\theta}_{\ssc 2}$ in (\ref{2-B}), i.e.:
\begin{equation}
\displaystyle S=t_{\ssc 2}-t_{\ssc 1}+\beta (\theta_{\ssc 2}\bar{\theta}_{\ssc 1}-\theta_{\ssc 1}\bar{\theta}_{\ssc 2}). \label{bbeta}
\end{equation}
The factor $\beta$ is just an inverse energy and this guarantees the dimensional consistency of Eq. (\ref{bbeta}).

Besides (\ref{2-B}), other intervals can be obtained as follows. Let us define \cite{undici} the left and right time variables as 
\begin{displaymath}
\displaystyle t_{\ssc L}\equiv t+\bar{\theta}\theta, \qquad\quad  t_{\ssc R}\equiv t-\bar{\theta}\theta. 
\end{displaymath}
They are complex variables. In fact, using the results of Appendix A, we have 
\begin{displaymath}
t_{\ssc L}^*=t_{\ssc R} \label{4-B}.
\end{displaymath}
It is then straightforward to prove that the following two distances are supersymmetric invariant
\begin{displaymath}
\begin{array}{l}
S_{\ssc L} \equiv t_{\ssc 2(L)}-t_{\ssc 1(R)}-2\bar{\theta}_{\ssc 1}\theta_{\ssc 2} \medskip\\
S_{\ssc R} \equiv t_{\ssc 2(R)}-t_{\ssc 1(L)}+2\bar{\theta}_{\ssc 2}\theta_{\ssc 1}. \label{5-B}
\end{array}
\end{displaymath}
It is also easy to find the following relations between $S$, $S_{\ssc L}$ and $S_{\ssc R}$:
\begin{displaymath}
\displaystyle S_{\ssc R}=S-\bar{\Delta}\Delta, \qquad \quad S_{\ssc L}=S+\bar{\Delta}\Delta,
\end{displaymath}
where $\Delta\equiv \theta_{\ssc 2}-\theta_{\ssc 1}$ and $\bar{\Delta}\equiv \bar{\theta}_{\ssc 2}-\bar{\theta}_{\ssc 1}$.
Note that, when $\Delta=0$, or when $\bar{\Delta}=0$, $S_{\ssc R}=S_{\ssc L}=S$. If both $\Delta$ and $\bar{\Delta}$ are zero, then the interval (\ref{2-B}) reduces to the usual time interval:
\begin{displaymath}
S=S_{\ssc R}=S_{\ssc  L}=t_{\ssc 2}-t_{\ssc 1}.
\end{displaymath}

\newpage

\section{Appendix }

In standard quantum mechanics an explicit dependence of the operators on time $t$ can be introduced by passing from the Schr\"odinger picture to the Heisenberg one via the following formula:
\begin{displaymath}
\displaystyle A_{\ssc H}(t)=e^{iHt/\hbar} A_{\ssc S} e^{-iHt/\hbar}.
\end{displaymath}
Let us notice that in the argument of the exponential the time parameter $t$ is coupled with the operator that generates the time evolution $H$.

As we said in Sec. 3, we can have an Heisenberg picture also with respect to $\theta$ and $\bar{\theta}$. The analog of the translation operator in $t$ are now the translations operators in $\theta$ and $\bar{\theta}$. So, for example, we can go to the Heisenberg picture of the field $\varphi^a$. In the paper we said that the result is 
\be
\exp \bigl[\theta Q + \bar{Q}\bar{\theta}\, \bigr] \,  \varphi^a \, \exp\bigl[-\theta Q-\bar{Q}\bar{\theta}\, \bigr]=\Phi^a(\theta,\bar{\theta}) \label{comm}
\ee
and we are now going to prove it.
Let us notice that $Q$ and $\bar{Q}$ are two commuting operators. Therefore the Baker-Campbell-Hausdorff formula reduces to:
\begin{displaymath}
\displaystyle
\exp \bigl[\theta Q+\bar{Q}\bar{\theta} \, \bigr]=\exp [\theta Q]\exp \bigl[\bar{Q}\bar{\theta} \, \bigr]=(1+\theta Q)(1+\bar{Q}\bar{\theta})
\end{displaymath}
and the LHS of (\ref{comm}) becomes:
\bea
&& \displaystyle e^{\theta Q+\bar{Q}\bar{\theta}}\varphi^a
e^{-\theta Q-\bar{Q}\bar{\theta}}=
(1+\theta Q)(1+\bar{Q}\bar{\theta}) \varphi^a
(1-\theta Q)(1-\bar{Q}\bar{\theta})=\nonumber\\
&& =\varphi^a+\theta \bigl[Q,\varphi^a\bigr]-\bar{\theta}\bigl[\bar{Q},\varphi^a\bigr]
+\bar{\theta}\theta \bigl[-\varphi^aQ\bar{Q}-\bar{Q}\varphi^a Q+Q\varphi^a\bar{Q}-Q\bar{Q}\varphi^a\bigr]=\nonumber\\
&& =\varphi^a+\theta c^a-\bar{\theta}\omega^{ba}\bar{c}_b+ \bar{\theta}\theta \bigl[[Q,\varphi^a],\bar{Q}\bigr]=\varphi^a+\theta c^a-\bar{\theta}\omega^{ba}\bar{c}_b+\bar{\theta}\theta[c^a,\bar{Q}]=\nonumber\\
&& =\varphi^a+\theta c^a+\bar{\theta}\omega^{ab}\bar{c}_b+i\bar{\theta}\theta \omega^{ab}\lambda_b= \nonumber\\
&& =\Phi^a.
\eea
This proves (\ref{comm}).
So we can look at the superfield operator $\Phi^a$ as the Heisenberg picture of the operator $\varphi^a$, with respect to the Grassmann partners of time $\theta$ and $\bar{\theta}$.

The procedure analyzed above can be applied also to  a generic function $F(\varphi)$. Let us define the operator:
\begin{displaymath}
S\equiv \exp\Bigl[\theta Q+\bar{Q}\bar{\theta}\Bigr]
\end{displaymath}
and let us see which type of transformation it induces on a generic function $F(\varphi)$. We will have:
\bea
\displaystyle SFS^{-1}&=&\exp \Bigl[\theta Q+\bar{Q}\bar{\theta}\Bigr]
F \exp \Bigl[-\theta Q-\bar{Q}\bar{\theta}\Bigr] =\nonumber\\
&=& F+\theta [Q,F]-\bar{\theta}[\bar{Q},F]+\bar{\theta}\theta\Bigl[[Q,F]\bar{Q}
+\bar{Q}[Q,F]\Bigr]=\nonumber\\
&=& F+\theta c^a\partial_a F-\bar{\theta}\bar{c}_a\omega^{ab}\partial_bF
+\bar{\theta}\theta \Bigl[\bar{Q},c^a\partial_aF\Bigr]=\nonumber\\
&=& F+\theta c^a\partial_a F-\bar{\theta}\bar{c}_a\omega^{ab}\partial_bF+\bar{\theta}\theta (-i \lambda_a\omega^{ab}\partial_bF+\bar{c}_a\omega^{ab}\partial_b\partial_dFc^d).
\nonumber
\eea
In the particular case in which the function $F$ is just the Hamiltonian $H$ we get that:
\be
\displaystyle SHS^{-1}=H+\theta N+\bar{N}\bar{\theta}-i\bar{\theta}\theta \widetilde{\cal H},
\label{mult}
\ee
i.e., the multiplet $SHS^{-1}$ includes both the Hamiltonian $H$ that appears in the weight of the quantum path integral and the Hamiltonian $\widetilde{\cal H}$ that appears in the weight of the classical path integral, besides the two conserved charges $N$ and $\bar{N}$. Eq. (\ref{mult}) can also be written in a compact form as:
\be
\displaystyle SH(\varphi)S^{-1}=H(\Phi), \label{mult2}
\ee
which means that the Heisenberg picture of the operator $H(\varphi)$ is just given by the same function $H$ but with the fields $\varphi$ replaced by the superfields $\Phi$. 

To prove that the RHS of (\ref{mult}) is just given by $H(\Phi)$ let us evaluate explicitly the expansion of $H(\Phi^a)=H(\varphi^a+ X^a)$ where $X^a \equiv \theta c^a +\bar{\theta}\omega^{ab}\bar{c}_b+i\bar{\theta}\theta\omega^{ab}\lambda_b$.
Using the properties of the Grassmann variables, the expansion of the Hamiltonian $H(\Phi)$ becomes:
\bea
\displaystyle H(\Phi)&=&H(\varphi)+X^a\partial_aH+\frac{1}{2} X^aX^b \partial_a\partial_bH=\nonumber\\
&=& H(\varphi)+ \theta c^a\partial_aH+\bar{\theta}\omega^{ab}\bar{c}_b\partial_aH +i\bar{\theta}\theta\omega^{ab}\lambda_b\partial_aH+\nonumber\\
&& +\frac{1}{2}(\theta c^a + \bar{\theta}\omega^{ae}\bar{c}_e) 
(\theta c^b+\bar{\theta}\omega^{bf}\bar{c}_f)\partial_a\partial_b H= \nonumber\\
&=&  H(\varphi)+\theta N -\bar{\theta}\bar{N} -i \bar{\theta}\theta(\lambda_a\omega^{ab}\partial_bH+i\bar{c}_a\omega^{ab}\partial_b\partial_dHc^d), \nonumber
\eea
which is just the RHS of Eq. (\ref{mult}). From Eq. (\ref{mult}) and (\ref{mult2}) we can derive that the relationship between $H$ and $\widetilde{\cal H}$ is given by:
\be
\displaystyle i \int d \theta d\bar{\theta} H(\Phi)=\widetilde{\cal H}. \label{cicinque}
\ee
A similar formula holds also for the Lagrangian. In particular, let us consider the Lagrangian 
\be
L=p\dot{q}-H(q,p). \label{lag}
\ee
If we replace the fields with the superfields in the kinetic term $p\dot{q}$ and we integrate over $\theta$, $\bar{\theta}$, what we get is
\bea
\displaystyle i \int d\theta d\bar{\theta} \;\Phi^p\dot{\Phi}^q &=&
i \int d \theta d\bar{\theta} (p+\theta c^p-\bar{\theta}\bar{c}_q-i\bar{\theta}\theta \lambda_q)(\dot{q}+\theta \dot{c}^q + \bar{\theta} \dot{\bar{c}}_p +i\bar{\theta}\theta \dot{\lambda}_p)=\nonumber \\
\displaystyle &=& \lambda_q \dot{q}+i\bar{c}_q\dot{c}^q+ic^p\dot{\bar{c}}_p-p\dot{\lambda}_p
=\lambda_a\dot{\varphi}^a+i\bar{c}_a\dot{c}^a-\frac{d}{dt}(\lambda_{p}p
+i\bar{c}_{p}c^{p}). \label{cisei}
\eea
Collecting together (\ref{cicinque}), (\ref{lag}) and (\ref{cisei}) we get exactly Eq. (\ref{33-1}):
\begin{displaymath}
\displaystyle i\int d\theta d\bar{\theta} L(\Phi)=\widetilde{\cal L}-
\frac{d}{dt}(\lambda_{p}p+i\bar{c}_{p}c^{p}).
\end{displaymath}

\newpage 

\section{Appendix }

In this Appendix we will give some details about the classical and the quantum path integrals in the momentum representation. First of all, we want to give some details on the derivation of Eq. (\ref{47-1}). Let us start analyzing the weight of the path integral. From (\ref{34-1}) we know that the weight appearing in the path integral of the kernel $\langle q,p, c^q,c^p,t|q_{\ssc 0},p_{\ssc 0},c^{q_0},c^{p_0},t_{\ssc 0}\rangle$ contains a Lagrangian $L[\Phi]$ plus some surface terms (\ref{34-2}) of the form:
\be
\displaystyle (\textrm{s.t.})=i\lambda_pp-i\lambda_{p_{\ssc 0}}p_{\ssc 0}-
\bar{c}_pc^p+\bar{c}_{p_{\ssc 0}}c^{p_{\ssc 0}}.  \label{diuno}
\ee 
When we insert the expression of the kernel (\ref{34-1}) into (\ref{46-1}) we see that we must consider, besides the terms of Eq. (\ref{diuno}), also the phase factors coming from the Fourier transforms (\ref{46-1}). These terms can be collected together and written in terms of the superfields as follows:
\begin{eqnarray}
&& \displaystyle \exp \Bigl[ i\lambda_pp-i\lambda_{p_{\ssc 0}}p_{\ssc 0}-
\bar{c}_pc^p+\bar{c}_{p_{\ssc 0}}c^{p_{\ssc 0}} -i\lambda_qq+\bar{c}_qc^q
+i\lambda_{q_0}q_{\ssc 0}-\bar{c}_{q_0}c^{q_0} \Bigr] =\nonumber\\
&& \displaystyle =\exp \left[- i\int id\theta d\bar{\theta} (q+\theta c^q+\bar{\theta} \bar{c}_p+i\bar{\theta}\theta \lambda_p)(p+\theta c^p-\bar{\theta} \bar{c}_q-i\bar{\theta}\theta \lambda_q) \right] \Biggl|_{t_0}^t = \nonumber \\
&& \displaystyle =\exp \left[-i \int_{t_0}^t i d\tau d\theta d\bar{\theta}\; \frac{d(QP)}{d\tau}\right].
\label{didue}
\end{eqnarray}
This proves that the surface terms appearing in the weight of Eq. (\ref{47-1}) are the right ones. 

If we want now to prove that, in going from (\ref{35-2}) to (\ref{47-1}), the functional measure changes from ${\mathscr D}^{\prime\prime}Q{\mathscr D}P$ to ${\mathscr D}Q{\mathscr D}^{\prime\prime}P$ then we must perform the same steps that are usually done in quantum mechanics to prove that the functional measure changes from ${\mathscr D}^{\prime\prime}q{\mathscr D}p$ to ${\mathscr D}q{\mathscr D}^{\prime\prime}p$ in passing from the path integral in the coordinate representation (\ref{38-1}) to the one in the momenta (\ref{49-3}). In particular, to understand the meaning of the functional measures ${\mathscr D}^{\prime\prime} q{\mathscr D}p$ and ${\mathscr D} q{\mathscr D}^{\prime\prime} p$, we will review the discretized versions of the quantum path integrals. Let us start with the one in the coordinate representation $\displaystyle \langle q,t|q_{\ssc 0},t_{\ssc 0}\rangle=\langle q| \Bigl[\exp -\frac{i}{\hbar}(t-t_{\ssc 0}) \widetilde{\cal H}\Bigr] |q_{\ssc 0}\rangle $.
If we slice the time interval $t-t_{\ssc 0}$ in $N$ intervals of length $\epsilon$ and we use the formula $\displaystyle \exp \left[-\frac{i}{\hbar}(t-t_{\ssc 0})\widetilde{\cal H}\right]= \exp \left(-\frac{i}{\hbar}\epsilon \widetilde{\cal H} \right)^N$ then we can insert a completeness relation in $q$ before every exponential factor and a completeness in $p$ after every exponential factor to get the following equation \cite{kleinert}:
\begin{eqnarray}
\displaystyle \langle q,t|q_{\ssc 0},t_{\ssc 0}\rangle&=&\int \frac{dp_{\ssc N}}{2\pi\hbar} dq_{\ssc N-1} \frac{dp_{\ssc N-1}}{2\pi\hbar}  \cdots dq_{\ssc 1}\frac{dp_{\ssc 1}}{2\pi\hbar} \langle q|\exp \left(-i\epsilon \widetilde{\cal H}/\hbar \right) |p_{\ssc N}\rangle \langle p_{\ssc N}|q_{\ssc N-1}\rangle \cdot \nonumber\\
&& \cdot \langle q_{\ssc N-1}| \exp \left(-i\epsilon \widetilde{\cal H}/\hbar \right)|p_{\ssc N-1}\rangle  \cdots \langle q_{\ssc 1}|\exp \left(-i\epsilon \widetilde{\cal H}/\hbar \right)|p_{\ssc 1}\rangle \langle p_{\ssc 1}|q_{\ssc 0}\rangle = \nonumber\\
&=&\lim_{N \to \infty} 
\prod_{n=1}^{N-1} \left[\int_{-\infty}^{\infty} dq_n\right]
\prod_{n=1}^{N}\left[\int_{-\infty}^{\infty}\frac{dp_n}{2\pi\hbar}\right] \exp\left[\frac{i}{\hbar}{\mathscr A}^{\ssc N}\right] \nonumber
\end{eqnarray}
with $\displaystyle {\mathscr A}^{\ssc N}=\sum_{n=1}^{N}\Bigl[ p_n(q_n-q_{n-1})-\epsilon H(p_n,q_n) \Bigr]$. In the previous path integral the initial coordinate $q_{\ssc 0}$ and the final one $q\equiv q_{\ssc N}$ are not integrated over, but they are fixed by the boundary conditions. All the momenta, from the initial $p_{\ssc 1}$ to the final $p_{\ssc N}$, are instead integrated over: this is the meaning of the functional measure ${\mathscr D}^{\prime\prime} q{\mathscr D}p$. Now, to go from the path integral in the coordinate representation to the one in the momenta, we must perform a Fourier transform over the initial and final coordinates:
\begin{displaymath}
\displaystyle \langle p,t|p_{\ssc 0},t_{\ssc 0}\rangle=\int dq\,dq_{\ssc 0} \, \exp \left[ -\frac{i}{\hbar}pq\right] \langle q, t|q_{\ssc 0},t_{\ssc 0}\rangle\, \exp \left[\frac{i}{\hbar}p_{\ssc 0}q_{\ssc 0}\right]. \label{momrep}
\end{displaymath}
Also in this case we can slice the time interval $t-t_{\ssc 0}$ into $N$ intervals of length $\epsilon$ and insert the completeness relations. This time we will insert the completeness relations in $q$ after each exponential factor $\displaystyle \exp \left(-i\epsilon \widetilde{\cal H}/\hbar\right)$ and those in $p$ before it. What we get is:
\begin{equation}
\begin{array}{l}
\displaystyle \int dq_{\ssc N} \frac{dp_{\ssc N-1}}{2\pi\hbar}dq_{\ssc N-1} \cdots \frac{dp_{\ssc 1}}{2\pi\hbar} dq_{\ssc 1}
\langle p| \exp \left( -i\epsilon \widetilde{\cal H}/\hbar \right) |q_{\ssc N}\rangle \langle q_{\ssc N} |p_{\ssc N-1}\rangle \langle p_{\ssc N-1}| 
\left(-i\epsilon \widetilde{\cal H}/\hbar \right)|q_{\ssc N-1}\rangle \cdots \label{51} \\
\displaystyle \cdots \langle p_{\ssc 1}| \exp\left( -i\epsilon \widetilde{\cal H}/\hbar \right)| q_{\ssc 1}\rangle \langle q_{\ssc 1}| p_{\ssc 0} \rangle=
\lim_{N \to \infty} 
\prod_{n=1}^{N} \left[\int_{-\infty}^{\infty} dq_n\right]
\prod_{n=1}^{N-1}\left[\int_{-\infty}^{\infty}\frac{dp_n}{2\pi\hbar}\right] \exp\left[\frac{i}{\hbar}{\mathscr B}^{\ssc N}\right].
\end{array}
\end{equation}
with $\displaystyle {\mathscr B}^{\ssc N}=\sum_{n=1}^N \left[ -q_{n}(p_{n}-p_{n-1}) 
-\epsilon H(p_n,q_{n})\right]$. 

In the path integral (\ref{51}) the initial and the final momenta $p_{\ssc 0}$ and $p\equiv p_{\ssc N}$ are fixed, while all the coordinates from $q_{\ssc 1}$ to $q_{\ssc N}$ are integrated over.
In the continuum limit Eq. (\ref{51}) reproduces the path integral (\ref{49-3}), i.e.:
\begin{displaymath}
\displaystyle \langle p,t|p_{\ssc 0},t_{\ssc 0}\rangle = \int {\mathscr D}q {\mathscr D}^{\prime\prime} p
\; \exp \left[ \frac{i}{\hbar} \int_{t_0}^{t} d\tau \Bigl[-q\dot{p}-H(q,p,t)\Bigr]\right]. \label{62}
\end{displaymath}
The final result of these manipulations in the continuum limit was to change the kinetic terms of the quantum path integral from $p\dot{q}$ to $-q\dot{p}$. These kinetic terms can be connected via the surface term $\displaystyle \frac{d}{dt}(qp)$, according to the formula: $\displaystyle -q\dot{p}= p\dot{q}-\frac{d}{dt}(qp)$. As we have seen in Eq. (\ref{didue}) this connection becomes 
$\displaystyle -Q\dot{P}=P\dot{Q}-\frac{d}{dt}(QP)$ at the classical path integral level.

\newpage 

\section{Appendix }

Up to now in this paper we have applied our dequantization procedure to the coordinate and the momentum representations of quantum mechanics. The procedure itself gave us the classical path integral in a particular representation. Starting from the eigenstates of the position\break $\hat{q}|q\rangle =q|q\rangle$ and replacing the fields with the superfields,  we got the relation: $ \hat{Q}|Q\rangle =Q |Q\rangle$, which allowed us to identify the states $|Q\rangle$ with the simultaneous eigenstates $|q,\lambda_p, c^q,\bar{c}_p\rangle$. Analogously, in the momentum representation the states $|P\rangle$ could be identified, via the relation $\hat{P}|P\rangle =P|P\rangle$, with the eigenstates $|P\rangle \equiv |p,\lambda_q,c^p,\bar{c}_q\rangle$. In order to write the path integral entirely in terms of the superfields we had to consider the kernel of transition between the states $|Q\rangle$ or $|P\rangle$ mentioned above.
In this Appendix we want to see what happens if we consider the coherent states representation.

First of all, let us remember that the coherent states in quantum mechanics are defined as the eigenstates $|z\rangle$ of the operator $\displaystyle \hat{a}=\frac{1}{\sqrt{2\hbar}}(\hat{q}+i\hat{p})$:
\be
\displaystyle \hat{a}|z\rangle =z |z\rangle.  \label{coh}
\ee
The transition amplitude between these states is given by \cite{klauder}:
\be
\displaystyle \langle z,t|z_{\ssc 0},t_{\ssc 0}\rangle = \lim_{N\to \infty} \pi^{\ssc -N} \int d^2z_{\ssc 1}\ldots d^2z_{\ssc N} \; \exp\Biggl\{ i \int_{t_0}^t d\tau\Biggl[ \frac{i}{2}\Bigl( z^*\frac{dz}{d\tau}-\frac{dz^*}{d\tau}z\Bigr)-\frac{1}{\hbar}H(z^*,z)\Biggr]\Biggr\}, \label{ampl}
\ee
where $\int dt H(z^*,z)$ is the continuum limit of $\sum_j \epsilon H(z_j^*,z_{j-1})$ with 
\begin{displaymath}
\displaystyle H(\alpha^*,\beta)\equiv\frac{\langle \alpha|H|\beta \rangle}{\langle \alpha|\beta \rangle}, \qquad |\alpha\rangle, |\beta \rangle \; {\textrm{coherent states}}.
\end{displaymath}
Since the operator $\hat{a}$ is not Hermitian, the variable $z$ is complex but it can be decomposed as 
\begin{displaymath}
\displaystyle z\equiv \frac{q+ip}{\sqrt{2\hbar}}. \label{reim0}
\end{displaymath}
One usually indicates the coherent state $|z\rangle$ with $|q,p\rangle$, see for example \cite{klauder}. Of course in this case $|q,p\rangle$ {\it does not} indicate the simultaneous eigenstate of the operators $\hat{q}$ and $\hat{p}$ with eigenvalues $q$ and $p$ which cannot exist in quantum mechanics as $\hat{q}$ and $\hat{p}$ do not commute. The parameters $q$ and $p$ within the ket $|\cdot , \cdot \rangle$ are simply the real and imaginary parts of the eigenvalue $z$ of the operator $\hat{a}$. 
In terms of the states $|q,p\rangle$ we can rewrite the path integral (\ref{ampl}) as:
\be
\displaystyle \langle q,p, t| q_{\ssc 0}, p_{\ssc 0}, t_{\ssc 0} \rangle = 
\int {\mathscr D}^{\prime\prime} q{\mathscr D}^{\prime\prime}p 
\exp 	\Biggl[\frac{i}{\hbar}\int_{t_{\ssc 0}}^{t} d\tau \biggl( -\frac{1}{2}q\dot{p}+\frac{1}{2}p\dot{q}-H\biggl(\frac{q-ip}{\sqrt{2\hbar}}, \frac{q+ip}{\sqrt{2\hbar}}\biggr)\biggr)\Biggl] \label{ampl2}
\ee
where in the last step the functional  measure is given by:
$\displaystyle {\mathscr D}^{\prime\prime}q{\mathscr D}^{\prime\prime}p=\lim_{N \to \infty} \prod_{j=1}^N \biggl(\int \frac{dq_jdp_j}{2\pi\hbar}\biggr)$.

Now let us build the classical path integral in the coherent states representation limiting ourselves to the bosonic part. Since the fundamental commutators that can be derived from the CPI are: $[\hat{q}, \hat{\lambda}_q]=[\hat{p},\hat{\lambda}_p]=i$, the operators defining the analog of the coherent states at the classical path integral level are:
\bea
&& \displaystyle \hat{a}_q\equiv \frac{1}{\sqrt{2}}\hat{q} +\frac{i}{\sqrt{2}}\hat{\lambda}_q, \qquad
\quad \hat{a}_q^{\dagger}\equiv \frac{1}{\sqrt{2}}\hat{q} -\frac{i}{\sqrt{2}}\hat{\lambda}_q, \nonumber\\
&& \displaystyle \hat{a}_p\equiv \frac{1}{\sqrt{2}}\hat{p} +\frac{i}{\sqrt{2}}\hat{\lambda}_p, \qquad
\quad \hat{a}_p^{\dagger}\equiv \frac{1}{\sqrt{2}}\hat{p} -\frac{i}{\sqrt{2}}\hat{\lambda}_p. 
\nonumber
\eea
It is easy to prove that the only commutators different from zero are the following ones:
\begin{displaymath}
\displaystyle [\hat{a}_q,\hat{a}_q^{\dagger}]=[\hat{a}_p,\hat{a}_p^{\dagger}]=1.
\end{displaymath}
The classical coherent states can be defined as the simultaneous eigenstates of the operators $\hat{a}_q$ and $\hat{a}_p$ with eigenvalues $z^q$ and $z^p$ respectively:
\be
\displaystyle \hat{a}_q|z^q,z^p\rangle =z^q|z^q,z^p\rangle, \qquad \hat{a}_p|z^q,z^p\rangle =z^p|z^q,z^p\rangle. \label{eigenvec}
\ee
The resolutions of the identity:
\be
\displaystyle \mathbb{I}=\frac{1}{\pi^2} \int \, d^2z^q d^2z^p \, |z^q,z^p\rangle \langle z^q,z^p |
\label{compl}
\ee
and the scalar product 
\be
\displaystyle \langle z^q_j,z^p_j | z^q_{j-1},z^p_{j-1} \rangle =\exp \biggl[ -\frac{1}{2} |z^q_j|^2-\frac{1}{2}|z^q_{j-1}|^2
\displaystyle +z^{q*}_jz^q_{j-1} -\frac{1}{2}|z^p_j|^2-\frac{1}{2}|z^p_{j-1}|^2+z^{p*}_jz^p_{j-1}\biggr] \label{scpr}
\ee
allow us to evaluate the classical path integral in the coherent states representation $(z^q,z^p)$. In fact, if we divide the interval $t-t_{\ssc 0}$ in $N+1$ intervals of length $\displaystyle \epsilon=\frac{t-t_{\ssc 0}}{N+1}$ and we use (\ref{compl})-(\ref{scpr}) we get the following discretized path integral:
\bea
&& \displaystyle \langle z^q, z^p,t|z^q_{\ssc 0},z^p_{\ssc 0},t_{\ssc 0} \rangle = \langle z^q,z^p| e^{-i \widetilde{\cal H} (t-t_{\ssc 0})}| z^q_{\ssc 0}, z^p_{\ssc 0}\rangle =\lim_{N \to \infty} \langle z^q, z^p| e^{ -i\epsilon \widetilde{\cal H}(N+1)} |z^q_{\ssc 0}, z^p_{\ssc 0} \rangle = \nonumber\\
&&= \displaystyle  \lim_{N \to \infty} \prod_{j=1}^{N} \biggl(\int \frac{d^2z^q_j}{\pi} \frac{d^2z^p_j}{\pi} \biggr) \; \prod_{j=1}^{N+1} \exp \biggl[ -\frac{1}{2} |z^q_j|^2-\frac{1}{2}|z^q_{j-1}|^2 +z^{q*}_jz^q_{j-1} \nonumber\\
&& -\frac{1}{2}|z^p_j|^2-\frac{1}{2}|z^p_{j-1}|^2+z^{p*}_jz^p_{j-1}\biggr]
\exp \biggl[-i\epsilon \widetilde{\cal H} (z^{q*}_j,z^{p*}_j, z^q_{j-1},z^p_{j-1})\biggr]
=\nonumber\\
&&=  \displaystyle  \lim_{N \to \infty} \prod_{j=1}^N \biggl( \int \frac{d^2z^q_j}{\pi}\frac{d^2z^p_j}{\pi} \biggr) \exp \biggl[ \sum_{j=1}^{N+1} \biggl\{ -\frac{1}{2} z^{q*}_j(z^q_j-z^q_{j-1})+\frac{1}{2} (z^{q*}_j-z^{q*}_{j-1})z^q_{j-1} \nonumber\\
&& \displaystyle -\frac{1}{2}z^{p*}_j(z^p_j-z^p_{j-1}) +\frac{1}{2}(z^{p*}_j-z^{p*}_{j-1})z^p_{j-1} -i\epsilon \widetilde{\cal H} (z^{q*}_j,z^{p*}_j,z^q_{j-1}, z^p_{j-1}) \biggr\} \biggr]. \label{cohpi}
\eea
In the continuum limit and in terms of the real and the imaginary parts of $z$:
\be
\displaystyle z^q\equiv \frac{q+i\lambda_q}{\sqrt{2}}, \qquad z^p\equiv \frac{p+i\lambda_p}{\sqrt{2}}, \label{reim}
\ee
the kernel of propagation (\ref{cohpi}) becomes:
\bea
\displaystyle && \langle q, \lambda_q,p,\lambda_p,t|q_{\ssc 0},\lambda_{q_{\ssc 0}},p_{\ssc 0}, \lambda_{p_{\ssc 0}}, t_{\ssc 0}\rangle =\int {\mathscr D}^{\prime\prime} \varphi {\mathscr D}^{\prime\prime} \lambda \;
\exp\biggl[ i \int_{t_{\ssc 0}}^{t} d\tau \biggl( \frac{1}{2}\lambda_q\dot{q} +\frac{1}{2}\lambda_p \dot{p} -\frac{1}{2}q\dot{\lambda}_q-\frac{1}{2}p\dot{\lambda}_p \nonumber\\
&& \qquad  -i\epsilon \widetilde{\cal H}\biggl(\frac{q_j-i\lambda_{q_j}}{\sqrt{2}},\frac{p_j-i\lambda_{p_j}}{\sqrt{2}},\frac{q_{j-1}+i\lambda_{q_{j-1}}}{\sqrt{2}},\frac{p_{j-1}+i\lambda_{p_{j-1}}}{\sqrt{2}}\biggr)\biggr)\biggl] \label{seg}
\eea
where $\displaystyle {\mathscr D}^{\prime\prime}\varphi {\mathscr D}^{\prime\prime}\lambda= \lim_{N\to \infty}  \prod_{j=1}^N \biggl(\int \frac{d q_j d\lambda_{q_j} dp_j d\lambda_{p_j}}{4\pi^2} \biggr)$.

Now, which is the connection between the quantum path integral (\ref{ampl2}) and the classical path integral (\ref{seg}) in the coherent state representations? If we replace the variables $q,p$ of (\ref{coh}) with the supervariables $Q$ and $P$ and do not consider the Grassmann part, Eq. (\ref{coh}) becomes:
\bea
\displaystyle \hat{\Phi}^a|Z\rangle &=& Z |Z\rangle \nonumber\\
&\Downarrow& \nonumber \\
\biggl[ \hat{q}+i\hat{p}+i\bar{\theta}\theta (\hat{\lambda}_p-i\hat{\lambda}_q) \biggr] |Z\rangle &=&
\biggl[ q+ip+i\bar{\theta}\theta(\lambda_p-i\lambda_q) \biggr] |Z \rangle. \nonumber
\eea
Differently from the coordinate and the momentum representation, we cannot identify $|Z\rangle$ with the simultaneous eigenstate of $\hat{\Phi}^a$ with eigenvalue $Z$. In fact, to get $q+ip$ as eigenvalue of $\hat{q}+i\hat{p}$ and $\lambda_p-i\lambda_q$ as eigenvalue of $\hat{\lambda}_p-i\hat{\lambda}_q$, we should consider the simultaneous eigenstate of $\hat{q}$, $\hat{p}$, $\hat{\lambda}_q$, $\hat{\lambda}_p$ which cannot exist since $\hat{\varphi}$ and $\hat{\lambda}$ do not commute. This immediately tells us that for the coherent states representation we must slightly change the dequantization rules at the level of the states of the theory. To find out the right dequantization rules, let us briefly review the definition of the coherent states for quantum mechanics given in \cite{klauder}. If $|0\rangle$ is the eigenstate of the operator $\hat{a}$ with eigenvalue $0$, then 
\be
|z \rangle \equiv \exp \Bigl[ z\hat{a}^{\dagger}-z^*\hat{a} \Bigr] |0\rangle \label{a20-1}
\ee
is just the eigenstate of the operator $\hat{a}$ with eigenvalue $z$, i.e. it satisfies Eq. (\ref{coh}). If we use the decomposition $\displaystyle z \equiv \frac{q+ip}{\sqrt{2\hbar}}$ we get that the operator entering the definition of the coherent states is:
\bea
\displaystyle \exp [z \hat{a}^{\dagger}-z^*\hat{a}] &=& \exp \biggl[\frac{1}{2\hbar}(q+ip)(\hat{q}-i\hat{p})-\frac{1}{2\hbar}(q-ip)(\hat{q}+i\hat{p})\biggr]=\nonumber\\
\displaystyle &=& \exp\Bigl[\frac{i}{\hbar}(p\hat{q}-q\hat{p})\Bigr] = U[q,p] \nonumber
\eea
and Eq. (\ref{a20-1}) can be rewritten as:
\be
\displaystyle |q,p\rangle \equiv \exp \Bigl[\frac{i}{\hbar}(p\hat{q}-q\hat{p})\Bigr]|0\rangle \,
\Longrightarrow \, |q,p\rangle =U[q,p] |0\rangle. \label{def}
\ee
It is just on the operator $U(q,p)$ of Eq. (\ref{def}) that we can apply our standard dequantization rules, i.e. replace the fields with the superfields:
\bea
&& \displaystyle \hat{q} \,  \rightarrow \, \hat{q}+\theta \hat{c}^q +\bar{\theta} \hat{\bar{c}}_p + i\bar{\theta}\theta \hat{\lambda}_p,
\qquad  q\, \rightarrow \, q+\theta c^q + \bar{\theta} \bar{c}_p + i\bar{\theta}\theta \lambda_p, \nonumber \medskip\\
&& \displaystyle \hat{p} \, \rightarrow \, \hat{p}+\theta \hat{c}^p-\bar{\theta}\hat{\bar{c}}_q-i\bar{\theta}\theta \hat{\lambda}_q, \qquad
p \, \rightarrow \, p+\theta c^p -\bar{\theta} \bar{c}_q -i\bar{\theta}\theta \lambda_q
\nonumber
\eea
and integrate over $\theta$ and $\bar{\theta}$. What we get is the following operator:
\bea
\displaystyle U[q,p] &\longrightarrow & U[Q,P]=\exp \Biggl[ i\int d\theta d\bar{\theta} \; i(p+\theta c^p-\bar{\theta}\bar{c}_q -i\bar{\theta}\theta\lambda_q)(\hat{q}+\theta \hat{c}^q +\bar{\theta}\hat{\bar{c}}_p+i\bar{\theta}\theta\hat{\lambda}_p) \nonumber\\
&& \displaystyle -i \int d\theta d\bar{\theta} \; i(q+\theta c^q + \bar{\theta} \bar{c}_p+i \bar{\theta}\theta \lambda_p)(\hat{p}+\theta \hat{c}^p-\bar{\theta}\hat{\bar{c}}_q-i\bar{\theta}\theta \hat{\lambda}_q)\Biggr]. \label{dtredici}
\eea
Let us use the operator (\ref{dtredici}) to define the following states: 
\bea
\displaystyle |Q, P \rangle & \equiv & U[Q,P] \, |0\rangle \nonumber\\
\displaystyle & \Downarrow & \label{defcoh}  \\
\displaystyle |Q,P\rangle \,\equiv \, \exp \Bigl[ i\lambda_q \hat{q}+i\lambda_p\hat{p}-iq\hat{\lambda}_q-ip\hat{\lambda}_p && \hspace{-1cm}
-c^q\hat{\bar{c}}_q-c^p\hat{\bar{c}}_p-\bar{c}_q\hat{c}^q-\bar{c}_p\hat{c}^p \Bigr]\, |0\rangle. \nonumber
\eea
Since the Hilbert space underlying the CPI can be split in the tensor product of a bosonic and a Grassmann subspace \cite{sette}, we can rewrite (\ref{defcoh}) as 
\bea
\displaystyle |Q,P\rangle & \equiv & \exp \Bigl[i\lambda_q \hat{q}+i\lambda_p \hat{p}-iq\hat{\lambda}_q -ip\hat{\lambda}_p \Bigr]|0\rangle_{\scriptscriptstyle B} \displaystyle \cdot \nonumber \\
&& \cdot \exp \Bigl[-c^q \hat{\bar{c}}_q -c^p\hat{\bar{c}}_p-\bar{c}_q\hat{c}^q -\bar{c}_p\hat{c}^p\Bigr] |0 \rangle_{\scriptscriptstyle F},  \label{poipoipoi}
\eea
where $|0\rangle_{\scriptscriptstyle B}$ is the eigenstate with eigenvalue 0 of the bosonic operators $\hat{a}_q$ and $\hat{a}_p$, while, as we will see later on, $|0\rangle_{\scriptscriptstyle F}$ is the simultaneous eigenstate of the Grassmann operators $\hat{\bar{c}}_q$ and $\hat{\bar{c}}_p$ with eigenvalue zero. Eq. (\ref{poipoipoi}) can be factorized as:
\begin{displaymath}
\left \{
\begin{array}{l}
\displaystyle  |q,\lambda_q,p,\lambda_p \rangle =\exp \Bigl[i\lambda_q\hat{q}+
i\lambda_p\hat{p}-iq\hat{\lambda}_q-ip\hat{\lambda}_p\Bigr] \, |0\rangle_{\scriptscriptstyle B}, \medskip \\
|c^q,\bar{c}_q,c^p,\bar{c}_p \rangle =\exp \Bigl[-c^q\hat{\bar{c}}_q-c^p\hat{\bar{c}}_p-\bar{c}_q\hat{c}^q -\bar{c}_p\hat{c}^p \Bigr]\, |0\rangle_{\scriptscriptstyle F}. \label{def2}
\end{array} \right.
\end{displaymath}
The states $|q,\lambda_q,p,\lambda_p\rangle$ are just the eigenstates of $\hat{a}_q$ and $\hat{a}_p$ with eigenvalues $z^q$ and $z^p$, i.e. they satisfy the equation (\ref{eigenvec}) with eigenvalues $z^q$ and $z^p$ given by Eq. (\ref{reim}).
As a consequence, if we limit ourselves to the bosonic part of the theory, it is easy to check that the rules to go from the quantum path integral (\ref{ampl2}) to the classical one (\ref{seg}) are the usual ones:

\noindent {\bf 1)} Replace in the QPI the variables $(q,p)$ with the supervariables 
\begin{displaymath}
Q=q+i\bar{\theta}\theta \lambda_p, \qquad \qquad P=p-i\bar{\theta}\theta \lambda_q.
\label{ttilde}
\end{displaymath}
{\bf 2)} Extend the time integration to the supertime integration multiplied by $\hbar$. 

Before concluding this section, let us briefly analyze what happens in the Grassmann part of the theory. It is easy to prove that, with the scalar product \cite{sette} under which $\hat{c}^{q\dagger}=\hat{\bar{c}}_q$ and $\hat{c}^{p\dagger}=\hat{\bar{c}}_p$, the operators $\hat{\bar{c}}_q$ and $\hat{\bar{c}}_p$ can be used to define the Grassmann coherent states. 
In fact, if we apply on the state $|0\rangle_{\ssc F}$ of Eq. (\ref{poipoipoi}) the operator
\begin{displaymath}
\hat{F}=\exp\Bigl[-c^q\hat{\bar{c}}_q-c^p\hat{\bar{c}}_p-\bar{c}_q\hat{c}^q-\bar{c}_p\hat{c}^p\Bigr]
\end{displaymath}
then it is easy to prove that the states 
\begin{displaymath}
\displaystyle \hat{F}|0\rangle_{\ssc F}=\exp\Bigl[-c^q\hat{\bar{c}}_q-c^p\hat{\bar{c}}_p-\bar{c}_q\hat{c}^q-\bar{c}_p\hat{c}^p
\Bigr] |0\rangle_{\ssc F}
\end{displaymath}
have just the properties of the Grassmannian odd coherent states outlined in \cite{klauder}, i.e. they are eigenstates of $\hat{\bar{c}}_q$ and $\hat{\bar{c}}_p$ with eigenvalues $\bar{c}_q$ and $\bar{c}_p$. 
So, for what concerns the Grassmann sector of the Hilbert space, the dequantization procedure via the superfields produces just the representation in which one diagonalizes the Grassmann operators $\hat{\bar{c}}_q$ and $\hat{\bar{c}}_p$. 
We can conclude this appendix by saying that the connection between the classical and the quantum path integrals via the superfield dequantization procedure works also in the case of coherent states, provided we write the coherent states of the CPI following Eq. (\ref{defcoh}):
\begin{displaymath}
\displaystyle |Q,P\rangle =\exp \Biggl[i \int i \,d\theta d\bar{\theta}(P\hat{Q}-Q\hat{P})\Biggr] |0\rangle.
\end{displaymath}

\newpage

\section{Appendix }

In this Appendix we want to prove that, like in standard quantum theory, it is possible to derive the classical path integral (\ref{38-2}) using, from the beginning, the completeness relations and the Trotter formula. Then we will write the graded commutators (\ref{14-2}), which characterize the operatorial theory associated to the CPI, in terms of the superfields.  

First of all let us remember that we can define the fundamental kets $|Q\rangle$ and $|P\rangle$ via the relations:
\begin{equation}
\left\{ \begin{array}{l}
\hat{Q}(\theta,\bar{\theta}) |Q\rangle = Q(\theta,\bar{\theta}) |Q\rangle \medskip \\
\hat{P}(\theta,\bar{\theta}) |P\rangle = P(\theta, \bar{\theta}) |P\rangle. \label{151}
\end{array} \right.
\end{equation}
From Eq. (\ref{151}) the fundamental kets $|Q\rangle$ and $|P\rangle$ are uniquely determined. They are the eigenstates of the operators $\hat{Q}(\theta,\bar{\theta})$ and $\hat{P}(\theta,\bar{\theta})$ with eigenvalues $Q(\theta,\bar{\theta})$ and $P(\theta,\bar{\theta})$. With this definition we can identify 
\begin{equation}
|Q\rangle \equiv |q,\lambda_p,c^q,\bar{c}_p\rangle,  \qquad \quad
|P\rangle \equiv |\lambda_q, p, \bar{c}_q, c^p \rangle. \label{152}
\end{equation}
If we assume that integrating over a superfield is equivalent to integrating over all its components then the completeness relations can be written in a compact form as:
\begin{displaymath}
\displaystyle \int dQ |Q\rangle \langle Q| = \mathbb{I}, 
\qquad \int dP |P\rangle \langle P| = \mathbb{I}, \label{153}
\end{displaymath}
where $dQ\equiv dqd\lambda_pdc^qd\bar{c}_p$ and $dP \equiv d\lambda_qdpd\bar{c}_qdc^p$. Now we can derive the CPI in terms of the superfields with the same techniques used in ordinary textbooks, such as the Trotter formula, the completeness relations, and so on. The starting point is given, as usual, by the operator $\displaystyle \exp \left[-i(t-t_{\ssc 0})\hat{\widetilde{\cal H}}\right]$ that can be rewritten as $\displaystyle \exp \left[ -i(t-t_{\ssc 0}) \int  i d\theta d\bar{\theta} \, H(\hat{\Phi})\right] $. To get the associated path integral we must sandwich the previous expression between $\langle Q, t|$ and $|Q_{\ssc 0}, t_{\ssc 0}\rangle$. If we divide the time interval $t-t_{\ssc 0}$ in $N$ intervals of length $\epsilon$ and we use the formula
\begin{displaymath}
\displaystyle \exp \left[ -i(t-t_{\ssc 0})\hat{\widetilde{\cal H}} \right] =\left( \exp \left[-i\epsilon \hat{\widetilde{\cal H}}\right]\right)^N
\end{displaymath}
then we get $N$ exponential factors $\exp \left[ -i\epsilon \hat{\widetilde{\cal H}} \right]$. Inserting a completeness relation in $Q$ before every exponential factor and a completeness relation in $P$ after it we easily obtain:
\begin{displaymath}
\displaystyle \langle Q,t|Q_{\ssc 0},t_{\ssc 0}\rangle =\lim_{N \to \infty} \int \prod_{j=1}^{N-1} dQ_j \prod_{j=1}^N dP_j \prod_{j=1}^N {\mathscr A}_{j,j-1},
\end{displaymath}
where ${\mathscr A}_{j,j-1}$ is given by
\begin{eqnarray}
\displaystyle {\mathscr A}_{j,j-1} &=& \langle P_j|Q_{j-1}\rangle \langle Q_j| \exp \left[-i\epsilon \hat{\widetilde{\cal H}}\right] |P_j\rangle= \nonumber\\
&=& \langle P_j|Q_{j-1} \rangle \langle Q_j| \exp \left[-i\epsilon \int i d\theta d\bar{\theta} H(\hat{Q},\hat{P})\right]|P_j\rangle. \nonumber
\end{eqnarray}
Using the definition (\ref{152}) of the states $|Q\rangle$ and $|P\rangle$ it is easy to prove that also the scalar product between them can be written in terms of the superfields: $\displaystyle \langle P|Q\rangle= \exp i\int id\theta d\bar{\theta} (-PQ)$.
Consequently we can rewrite ${\mathscr A}_{j,j-1}$ as:
\begin{eqnarray}
\displaystyle {\mathscr A}_{j,j-1} &=&
\exp \left[ i \int i d\theta d\bar{\theta} \left( -P_jQ_{j-1}+P_jQ_j-\epsilon H[\Phi_j]\right)\right] = \nonumber\\
&=& \exp \left[ i\epsilon \int i d\theta d\bar{\theta} \left(P_j\frac{Q_j-Q_{j-1}}{\epsilon}-H[\Phi_j]\right)\right]. \nonumber
\end{eqnarray}
In the limit $N \to \infty$ and $\epsilon \to 0$ we get just the classical path integral (\ref{38-2}) written in terms of the superfields:
\begin{equation}
\displaystyle \langle Q,t|Q_{\ssc 0},t_{\ssc 0}\rangle =\int {\mathscr D}^{\prime\prime}Q(\theta,\bar{\theta}){\mathscr D}P(\theta,\bar{\theta}) \exp \left[ i\int_{t_0}^t id\tau d\theta d\bar{\theta} \left(P(\theta,\bar{\theta})\dot{Q}(\theta,\bar{\theta})-H(\Phi(\theta,\bar{\theta})) \right) \right]. \label{ccppii}
\end{equation}

As we said before, from the previous expression it is possible to derive the graded commutators (\ref{14-2}) and to write them in terms of the superfields. 
To do this, we can use an analogy between the CPI and the usual quantum {\it field} theories. For example the quantum phase space path integral for a scalar field $\phi({\bf x})$ is given by:
\begin{displaymath}
\displaystyle \int {\mathscr D}^{\prime\prime} \phi({\bf x}) {\mathscr D}\pi({\bf x}) \; \exp \frac{i}{\hbar} \int dtd {\bf x} \, \left[ \pi({\bf x})\dot{\phi}({\bf x})-H({\bf x})\right]
\end{displaymath}
and it is equivalent to an operatorial theory in which the only non-zero equal time commutators are:
\begin{equation}
\displaystyle \left[\phi({\bf x},t),\pi({\bf y},t)\right] = i\hbar \delta({\bf x}-{\bf y}).
\label{fseibis}
\end{equation}
Which is the operatorial theory lying behind the classical path integral (\ref{ccppii})? If we look at $Q(\theta,\bar{\theta})$ and $P(\theta,\bar{\theta})$ as fields in which the space variables ${\bf x}$ are replaced by the Grassmann variables $\theta$ and $\bar{\theta}$, we expect that the equal time commutator satisfied by $Q$ and $P$, analogous to (\ref{fseibis}), is the following one:
\begin{equation}
\displaystyle \left[ Q(t,\theta,\bar{\theta}), P(t,\theta^{\prime}, \bar{\theta}^{\prime})\right]=
\delta(\bar{\theta}-\bar{\theta}^{\prime})\delta(\theta-\theta^{\prime}). \label{121}
\end{equation}
Let us see whether, using the definition of the superfields 
\begin{eqnarray}
&& Q(t,\theta,\bar{\theta})=q(t) + \theta c^q(t) +\bar{\theta} \bar{c}_p(t)+ i\bar{\theta}\theta \lambda_p(t) \bigskip \nonumber \\
&& P(t,\theta^{\prime}, \bar{\theta}^{\prime})=p(t)+\theta^{\prime}c^p(t)-\bar{\theta}^{\prime}\bar{c}_q(t)-i\bar{\theta}^{\prime}\theta^{\prime}\lambda_q(t),
\nonumber 
\end{eqnarray}
Eq. (\ref{121}) implies just the commutators (\ref{14-2}) of the CPI. Because of the properties of the Grassmannian Dirac deltas,  Eq. (\ref{121}) can be rewritten as
\begin{equation}
\displaystyle \left[ Q(\theta,\bar{\theta}), P(\theta^{\prime},\bar{\theta}^{\prime}) \right] = \bar{\theta}\theta -\bar{\theta}\theta^{\prime}-\bar{\theta}^{\prime}\theta +\bar{\theta}^{\prime}\theta^{\prime}. \label{132}
\end{equation}
Expanding the LHS of (\ref{132}) in terms of $\theta$, $\bar{\theta}$, $\theta^{\prime}$ and $\bar{\theta}^{\prime}$ we get:
\begin{equation}
\displaystyle \left[ Q(\theta,\bar{\theta}), P(\theta^{\prime},\bar{\theta}^{\prime}) \right] = -i\bar{\theta}^{\prime}\theta^{\prime} \, [q,\lambda_q] +\theta \bar{\theta}^{\prime} \, [c^q, \bar{c}_q]-\bar{\theta}\theta^{\prime} \, [\bar{c}_p,c^p]
+i\bar{\theta}\theta \, [\lambda_p,p]+ \cdots.  \label{131}
\end{equation}
Comparing (\ref{132}) and (\ref{131}) we obtain that the non-zero graded commutators are:
\begin{displaymath}
\displaystyle [q,\lambda_q]=i, \qquad [c^q,\bar{c}_q]=1, \qquad [\bar{c}_p,c^p]=1, \qquad
[\lambda_p,p]=-i,
\end{displaymath}
which are precisely the graded commutators (\ref{14-2}) of the CPI. Of course, since on the LHS of (\ref{131}) we have a commutator and not an anticommutator, we get that:
\begin{displaymath}
\displaystyle \left[ P(t,\theta,\bar{\theta}), Q(t,\theta^{\prime},\bar{\theta}^{\prime}) \right]
=-\delta(\bar{\theta}-\bar{\theta}^{\prime})\delta(\theta-\theta^{\prime}),
\end{displaymath}
while $[Q,Q]=[P,P]=0$. Summarizing, we can say that the graded commutators of the CPI can be written in a compact form, using the superfields, as:
\begin{displaymath}
\displaystyle \left[ \Phi^a(t,\theta,\bar{\theta}),\Phi^b(t,\theta^{\prime},\bar{\theta}^{\prime}) \right]=\omega^{ab} \delta(\bar{\theta}-\bar{\theta}^{\prime})\delta(\theta-\theta^{\prime}).
\end{displaymath}
\newpage

\section{Appendix }

In this Appendix we want to analyze the issue of ordering problems in the CPI. In quantum mechanics the ordering problems arise because $\hat{q}$ and $\hat{p}$ do not commute, so there is more than one Hermitian operator $H(\hat{q},\hat{p})$ associated with the same classical Hamiltonian $H(q,p)$. For this reason we must specify the order in which we consider the operators $\hat{q}$ and $\hat{p}$ within the Hamiltonian $H(\hat{q},\hat{p})$. At the path integral level the different orderings correspond to different possible discretizations.  Before analyzing what happens in the CPI, let us notice that also in quantum mechanics no ordering problem arises if we consider Hamiltonians of the form $H(q,p)=p^2/2+V(q)$. There are instead problems when we couple $q$ and $p$ within the argument of the Hamiltonian $H$. For example
\begin{equation}
\displaystyle \hat{p}^2\hat{q}^2+\hat{q}^2\hat{p}^2+\hat{q}\hat{p}^2\hat{q}, \qquad \quad  \hat{p}\hat{q}\hat{p}\hat{q}+\hat{q}\hat{p}\hat{q}\hat{p}+\hat{p}\hat{q}^2\hat{p}
\label{ordpro}
\end{equation}
are two different Hermitian operators associated with the same classical observable $3q^2p^2$. Using the fundamental commutator $[\hat{q},\hat{p}]=i\hbar$ it is quite easy to prove that 
\begin{displaymath}
\displaystyle \hat{p}^2\hat{q}^2+\hat{q}^2\hat{p}^2+\hat{q}\hat{p}^2\hat{q}
=\hat{p}\hat{q}\hat{p}\hat{q}+\hat{q}\hat{p}\hat{q}\hat{p}+\hat{p}\hat{q}^2\hat{p}-\hbar^2,
\end{displaymath} 
which implies that, even if they are both Hermitian and associated with the same classical observable, the two operators of Eq. (\ref{ordpro}) are not equivalent but they differ for $\hbar^2$ terms. 

In the CPI $\hat{q}$ and $\hat{p}$ commute. Nevertheless $\hat{\varphi}^a$ and $\hat{\lambda}_a$ do not commute and one should analyze whether there are or not ordering problems.
Let us limit ourselves to the bosonic part of the theory. The bosonic part of the evolution operator appearing in the weight of the CPI is the Liouvillian $\hat{L}=\lambda_a\omega^{ab}\partial_bH$. If we consider a Hamiltonian $H(q,p)=q^np^m$, then the terms of the Liouvillian can be ordered in one of the following different ways:
\begin{eqnarray}
\displaystyle \hat{L} &=& m \left[\alpha_{\ssc 1}\lambda_q q^n p^{m-1}+\alpha_{\ssc 2}q\lambda_qq^{n-1}p^{m-1}+ \ldots +\alpha_{\ssc n+1} \,q^n\lambda_qp^{m-1} \right]+\nonumber \bigskip \\
&& -n \left[ \beta_{\ssc 1}\lambda_pp^mq^{n-1}+\beta_{\ssc 2}p\lambda_pp^{m-1}q^{n-1}+\ldots + \beta_{\ssc m+1}\,p^m\lambda_pq^{n-1} \right], \label{161}
\end{eqnarray}
where the sum of the weights $\alpha_j$ and $\beta_j$ is normalized as $\displaystyle \sum_{j=1}^{n+1}\alpha_j
=\sum_{j=1}^{m+1}\beta_j=1$. If we calculate the Hermitian conjugate of (\ref{161}) then we get:
\begin{eqnarray}
\displaystyle \hat{L}^{\dagger} &=& m\left[ \alpha_{\ssc 1}q^n\lambda_qp^{m-1}+\alpha_{\ssc 2} q^{n-1}\lambda_qqp^{m-1}+\ldots + \alpha_{\ssc n+1}\lambda_qq^np^{m-1} \right] +\nonumber \bigskip \\
&& - n\left[ \beta_{\ssc 1}p^m\lambda_pq^{n-1}+\beta_{\ssc 2}p^{m-1}\lambda_ppq^{n-1}+\ldots + \beta_{\ssc m+1}\lambda_pp^mq^{n-1} \right] \label{162}
\end{eqnarray}
where we have used the fact that with the standard scalar product, $\langle \psi|\tau \rangle = \int d\varphi \,\psi^*(\varphi) \tau(\varphi)$, both $\hat{\varphi}$ and $\hat{\lambda}$ are Hermitian operators. Since the ordering (\ref{161}) has to guarantee the Hermiticity of the Liouvillian $\hat{L}$, we must impose that $\hat{L}^{\dagger}-\hat{L}=0$, which is equivalent, using (\ref{161}) and (\ref{162}), to the following equation among the coefficients $\alpha_j$ and $\beta_j$:
\begin{equation}
\displaystyle m \sum_{j=1}^{n+1} \alpha_j [n-2(j-1) ]=n \sum_{j=1}^{m+1}\beta_j [m-2(j-1)] \label{181}.
\end{equation}
If the coefficients $\alpha_j$ and $\beta_j$ satisfy the previous equation then the Liouvillian (\ref{161}) is Hermitian with the associated ordering. An example of Hermitian ordering is the one for which the coefficients satisfy: $\alpha_j=\alpha_{n-j+2}$ and $\beta_j=\beta_{m-j+2}$. This ordering generalizes the Weyl one in which all the coefficients $\alpha$ and $\beta$ are equal: $\alpha_j=\frac{1}{n+1}$ and $\beta_j=\frac{1}{m+1}$. The pre-point ordering $\alpha_j=\delta_{j,1}$ and $\beta_j=\delta_{j,1}$ satisfies Eq. (\ref{181}) and corresponds to the Liouvillian in which all the operators $\lambda$ are on the right 
of the operators $\varphi$, i.e. $\hat{L}=\lambda_a\omega^{ab}\partial_bH$. The end-point ordering corresponds instead to the following choice of the coefficients: $\alpha_j=\delta_{j,n+1}$ and $\beta_j=\delta_{j,m+1}$ that satisfies (\ref{181}) and produces the Liouvillian $\hat{L}=\omega^{ab}\partial_bH \lambda_a$ with all the operators $\lambda$ on the right of the operators $\varphi$.

Before going on, let us notice that, besides Eq. (\ref{181}), which guarantees the Hermiticity of the Liouvillian, the coefficients $\alpha_j$ and $\beta_j$ must satisfy also the condition of normalization: $\sum_j\alpha_j=\sum_j \beta_j=1$ that can also be written as 
\begin{equation}
\displaystyle mn \sum_{j=1}^{n+1}\alpha_j =mn \sum_{j=1}^{m+1}\beta_j. \label{191}
\end{equation}
Making the difference between (\ref{191}) and (\ref{181}) we get:
\begin{equation}
\displaystyle m\sum_{j=1}^{n+1}\alpha_j (j-1)=n\sum_{j=1}^{m+1}\beta_j(j-1). \label{192}
\end{equation}
The previous condition is crucial for proving that, even if there is more than one possible Hermitian Liouvillian, all the Hermitian Liouvillians are equivalent to the pre-point one\break
$\hat{L}=\lambda_a\omega^{ab}\partial_bH$. In fact let us rewrite the Liouvillian $\hat{L}$ in a compact form as:
\begin{equation}
\displaystyle \hat{L}=m\sum_{j=1}^{n+1}\alpha_jq^{j-1}\lambda_qq^{n-j+1}p^{m-1}
-n\sum_{j=1}^{m+1} \beta_jp^{j-1}\lambda_pp^{m-j+1}q^{n-1}. \label{194}
\end{equation}
If we use the commutator $[\varphi^a,\lambda_b]=i\delta_b^a$ then we can rewrite (\ref{194}) as
\begin{eqnarray}
\displaystyle \hat{L}&=&m\sum_{j=1}^{n+1}\alpha_j\lambda_qq^np^{m-1}+m\sum_{j=1}^{n+1}i \alpha_j(j-1) q^{j-2+n-j+1}p^{m-1} +\nonumber\\
&& -n \sum_{j=1}^{m+1}\beta_j\lambda_pp^mq^{n-1}-n\sum_{j=1}^{m+1} i\beta_j(j-1) p^{j-2+m-j+1}q^{n-1}=\nonumber \\
&=&m\lambda_qq^np^{m-1}-n\lambda_pp^mq^{n-1}+iq^{n-1}p^{m-1} \left( m\sum_{j=1}^{n+1}\alpha_j(j-1) -n\sum_{j=1}^{m+1}\beta_j(j-1)\right). \nonumber
\end{eqnarray}
Using Eq. (\ref{192}) it turns out the last term above is zero, so we can conclude that the most general Hermitian Liouvillian $\hat{L}$ associated to the Hamiltonian $H=q^np^m$ is equivalent to the pre-point Liouvillian in which all the operators $\lambda$ are on the left of the operators $\varphi$. This proves that all the Liouvillians are equivalent and that there is no ordering problem for what concerns the bosonic part of the CPI. 

The discussion regarding the Grassmann part of the CPI is more complicated, due to the fact that there is more than one possible scalar product, as explained in the fourth of Refs. \cite{sette}. Without entering into further details we can say that the only scalar products that assure the Hermiticity of the Grassmann part of the Hamiltonian $\widetilde{\cal H}$ independently from the particular physical system that we consider, are the gauge and the symplectic scalar products defined in \cite{sette}. In these cases there are only two possible orderings, which guarantee the Hermiticity of the Hamiltonian: 
\begin{displaymath}
\widetilde{\cal H}_{\ssc \textrm{G,1}}=i\bar{c}_a\omega^{ab}\partial_b\partial_dH c^d, \qquad 
\widetilde{\cal H}_{\ssc \textrm{G,2}}=-ic^d\omega^{ab}\partial_b\partial_dH\bar{c}_a.
\end{displaymath}
They are completely equivalent, as we can easily prove using the graded commutators of the theory:
\begin{eqnarray}
\displaystyle -ic^d\omega^{ab}\partial_b\partial_dH\bar{c}_a &=& i\bar{c}_a\omega^{ab}\partial_b\partial_dHc^d -i\omega^{ab}\partial_b\partial_dH \bigl[ \bar{c}_a, c^d \bigr]= \nonumber\\
&=& i\bar{c}_a\omega^{ab}\partial_b\partial_dHc^d -i\omega^{db}\partial_b\partial_dH=
i\bar{c}_a\omega^{ab}\partial_b\partial_dHc^d. \nonumber
\end{eqnarray}
So we can conclude that also for what concerns the Grassmann part of the Hamiltonian $\widetilde{\cal H}$ there is more than one possible Hermitian ordering but also in this case the associated Hamiltonians are equivalent and no ordering problem arises. 

\newpage

\section{Appendix }

We want to give here a further ``rough" proof of (\ref{77-1}).
Let us consider a generic susy transformation of the CPI:
\begin{eqnarray}
&& t^{\prime}=t-\epsilon \,\bar{\theta}+\bar{\epsilon}\,\theta \nonumber\\
&& \theta^{\prime}=\theta-\epsilon \label{susy} \\
&& \bar{\theta}^{\prime}=\bar{\theta}+\bar{\epsilon}. \nonumber
\end{eqnarray}
The superdeterminant of the transformation $(t,\theta,\bar{\theta}) \to (t^{\prime}, \theta^{\prime},\bar{\theta}^{\prime})$ is equal to one:
\begin{displaymath}
\displaystyle J=\textrm{sdet} \begin{pmatrix}1 & -\epsilon & \bar{\epsilon}\\ 0 & 1 & 0\\ 0 & 0 & 1\end{pmatrix}=1. 
\end{displaymath}
So under the susy transformation (\ref{susy}) the integration measure $dtd\theta d\bar{\theta}$ is invariant:
\begin{equation}
\displaystyle dtd\theta d\bar{\theta}=dt^{\prime}d\theta^{\prime} d\bar{\theta}^{\,\prime}
\; \Longrightarrow \; \frac{dt}{dt^{\prime}}=\frac{d\theta^{\prime}}{d\theta}
\frac{d\bar{\theta}^{\,\prime}}{d\bar{\theta}}. \label{Idue}
\end{equation}
Since $\displaystyle \int d\theta \,\theta=\int d\theta^{\prime}\,\theta^{\prime}=1$ we will have that $\displaystyle \frac{d\theta^{\prime}}{d\theta}=\frac{\theta}{\theta^{\prime}}$.
Analogously from $\displaystyle \int d\bar{\theta}\,\bar{\theta}=\int d\bar{\theta}^{\prime}\,\bar{\theta}^{\prime}=1$ we can derive that $\displaystyle \frac{d\bar{\theta}^{\,\prime}}{d\bar{\theta}}=\frac{\bar{\theta}}{\bar{\theta}^{\prime}}$.
So, from (\ref{Idue}) and what we have derived above, we get
\begin{displaymath}
\displaystyle \frac{dt}{dt^{\prime}}=\frac{d\theta^{\prime}}{d\theta}\frac{d\bar{\theta}^{\,\prime}}{d\bar{\theta}}=\frac{\theta \bar{\theta}}{\theta^{\prime}\bar{\theta}^{\prime}}=\frac{\theta\bar{\theta}}{(\theta\bar{\theta})^{\prime}}
\end{displaymath}
from which we can ``somehow" identify $dt \sim \theta\bar{\theta}$ and $dt^{\prime}\sim (\theta\bar{\theta})^{\prime}$.


\end{document}